\documentclass[fleqn,usenatbib]{mnras}
\usepackage{savesym}
\usepackage{amsmath}	
\savesymbol{iint}
\usepackage{txfonts}
\usepackage[T1]{fontenc}
\usepackage{ae,aecompl}
\usepackage{graphicx}	
\usepackage{amssymb}	
\usepackage{enumerate}   
\usepackage{booktabs}   
\usepackage{float}
\usepackage{xcolor} 
\usepackage{natbib}

\newcommand{\unit}[1]{\ensuremath{\, \mathrm{#1}}}  

\title[Super-Earths' chemical inventory]{Chemical Fingerprints of Formation in Rocky Super-Earths' Data}

\author[Plotnykov \& Valencia]{
Mykhaylo Plotnykov,$^{1}$
Diana Valencia$^{2,3}$
\\
$^{1}$Department of Physics, University of Toronto, Toronto, ON M5S 3H4, Canada\\
$^{2}$Department of Physical \& Environmental Sciences, University of Toronto at Scarborough, Toronto, ON M1C 1A4, Canada\\
$^{3}$Department of Astronomy \& Astrophysics, University of Toronto, Toronto, ON M5S 3H4, Canada\\
}

\date{Accepted XXX. Received YYY; in original form ZZZ}

\pubyear{2020}

\begin{document}
\label{firstpage}
\pagerange{\pageref{firstpage}--\pageref{lastpage}}
\maketitle
\begin{abstract}
The composition of rocky exoplanets in the context of stars' composition provides important constraints to formation theories. 
In this study, we select a sample of exoplanets with mass and radius measurements with an uncertainty $<25\%$ and obtain their interior structure.
We calculate compositional markers, ratios of iron to magnesium and silicon, as well as core-mass fractions (cmf) that fit the planetary parameters, and compare them to the stars'.
We find four key results that successful planet formation theories need to predict: (1) In a population sense,  the composition of rocky planets spans a wider range than stars. 
The stars' Fe/Si distribution is close to a Gaussian distribution $1.63^{+0.91}_{-0.85}$, while the planets' distribution peaks at lower values and has a longer tail, $1.15^{+1.43}_{-0.76}$. 
It is easier to see the discrepancy in cmf space, where primordial stellar composition is $0.32^{+0.14}_{-0.12}$, while rocky planets' follow a broader distribution  $0.24^{+0.33}_{-0.18}$. 
(2) We introduce uncompressed density ($\overline{\rho_0}$ at reference pressure/temperature) as a metric to compare compositions. 
With this, we find what seems to be the maximum iron enrichment that rocky planets attain during formation ($\overline{\rho_0}\sim 6$  and cmf $\sim 0.8$). 
(3) Highly irradiated planets exhibit a large range of compositions.
If these planets are the result of atmospheric evaporation, iron enrichment and perhaps depletion must happen before gas dispersal. 
And (4), we identify a group of highly-irradiated planets that, if rocky, would be 2-fold depleted in Fe/Si with respect to the stars. 
Without a reliable theory for forming iron-depleted planets, these are interesting targets for follow up. 
\end{abstract}



\section{Introduction}
    Past and ongoing observational efforts have discovered an abundance of exoplanets so far, of which thousands are super-Earths and/or mini-Neptunes.
    Within these low-mass exoplanets, 60 of them have measured masses and radii that allow for compositional inference.
    Determining the composition of these planets is paramount to understanding how planets form, as it provides another axis of information to constrain formation theories. 
    However, determining composition is problematic for low-mass exoplanets because of two reasons.
    The practical reason is that error estimates in radius, and especially mass are large and thus, do not constrain composition precisely.
    
    The second, more fundamental reason, is that composition of low-mass exoplanets is plagued with degeneracies (\citealt{Valencia2007, Rogers2010, Zeng2013}). 
    With four compositional building blocks --- H/He, water/ices (either in solid, liquid or gaseous form), silicate mantles, and/or iron cores  --- and only two measurements: mass and radius, the problem is under-constrained. 
    The main degeneracy arises from trade-offs between the different building blocks, but other degeneracies are also in place. 
    For example, mini-Neptunes can efficiently substitute portions of water/ice layer with rocky content to yield the same bulk density, but also have trade-off between opacities and hydrogen-helium content \citep{Valencia2013}, or even ohmic dissipation and H/He content \citep{Pu2017} if in the right temperature regime.
    Thus, even with accurate mass and radius measurements, there are numerous compositional solutions.  
    
    In response to this, a few studies (e.g. \citealt{Dorn2015, Santos2015, Brugger2017}) have suggested that a way to break the degeneracy is to use the host star's refractory composition to constrain the refractory content of planets. 
    If indeed the refractory content of the planets is the same as of the host star, we could estimate the mass ratio of mantle to core, and reduce the degeneracy by one degree. 
    We will see in this study by looking at the data that this assumption is called into question. 
    
    On the other hand, if we restrict ourselves to planets that are suspected to be rocky, measurements of mass and radius constrain the core mass fraction (cmf) somewhat uniquely.
    In fact, in this study we show that with the current precision, mass and radius data constrain the iron to magnesium (Fe/Mg) or iron to silicate (Fe/Si) contents,  irrespective of the planet's degree of differentiation (i.e. amount of iron partitioned into the core versus mantle). 
    Consequently, we can compare the refractory ratios of rocky planets to those of the stars, and quantify the differences. 
    This may allow us to uncover the signature of planet formation in the composition of planets. 
    
    In practice, however, it is not possible to know a-priori which planets are undoubtedly rocky from mass and radius alone, precisely because of compositional degeneracy.
    Therefore, unless atmospheric characterization has deemed the absence of an atmosphere, such as the case for the recently observed LHS 3844b \citep{Kreidberg2019}, low-mass exoplanets can only be categorized as 'likely rocky' if there are smaller than the maximum size a rock can have, which we term the Rocky Threshold Radius (RTR), or certainly volatile rich, if larger.
    Also, according to core accretion theories, the more compact the planet is, the more likely it is to be rocky.  
    That is, it is more difficult to envision a scenario where the planet is made of only iron and H/He. 
    
    Furthermore, there is a radius gap \citep{Fulton2017} in the exoplanet population around FGK stars that seems consistent with the small-sized planets being the product of substantial atmospheric evaporation \citep{Owen2017}, while the larger planets have retained their envelopes.
    Therefore, it is possible that many of the super-Earths that are smaller than the RTR are indeed rocky. 
    If not, by assuming they are,  we infer the minimum cmf (or Fe/Mg content).  
    That is, if the planet had a larger cmf, being that iron is denser, the rocky portion (mantle + core) would be more compact and thus, one would need to invoke an envelope for it to have the same radius. 
    
    With this in mind, in this study we systematically constrain the cmf, Fe/Mg, and Fe/Si ratios for all the low-mass exoplanets for which there is well constrained observational data for mass and radius (i.e.  $ \leq 25\%$ respectively). 
    We compare these planetary refractory ratios to that of the stars as a population, as well as directly for four systems where the host star's composition is measured.
    We foresee this work as the first step into building a large database that will enable systematic comparisons between the composition of planets and stars. 
    
\section{Interior Structure Model}
\subsection{Structure of Super-Earth}

    To obtain the composition of the rocky exoplanets we use the interior structure model \texttt{SuperEarth} developed by \citealt{Valencia2006, Valencia2007}.
    Here, we summarize the main components, as well as explain the updates in composition that we implemented to carry out this work. 
    
    Planets are divided into three main layers consisting of core, mantle and water/ices, and such divisions are set by mass ratios. 
    For the purposes of this study we have ignored any water layer, as all but one planet investigated are too hot to have a liquid/solid water on their surfaces. 
    For the case of LHS~1140b, there is a possibility of liquid/solid H$_2$O content. 
    However, by assuming all planets are rocky,  we obtain the  minimum values for cmfs. 
    Where planets are assumed to be composed mainly of magnesium (Mg), silicon (Si), iron (Fe) and oxygen (O), while other minor and trace elements are ignored ( i.e aluminium (Al), calcium (Ca), etc.).  
    We assume all mantle minerals have the same iron number (Fe\# = $\mathrm{\frac{Fe}{Mg+Fe}}$). 
    Below, we quantify the effects of this assumption and show it is adequate for our purposes.

    \begin{figure}
        \includegraphics[width=\linewidth]{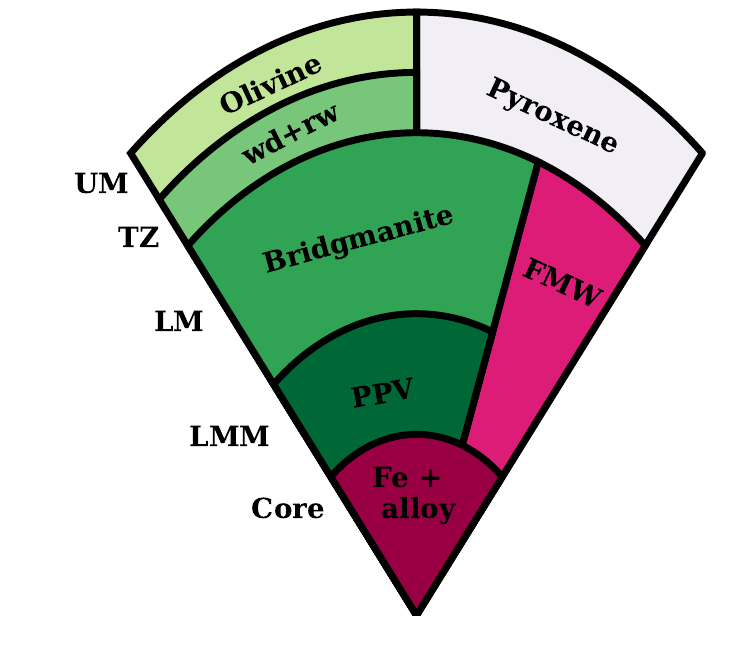}
        \caption{Cross-section of the interior structure of rocky planets considered in our model. 
        The mantle is composed of four layers: the upper mantle (UM) is composed of olivine and pyroxene minerals in different proportions ($50\%$ shown here), the transition zone (TZ) combines wadsleyite (wd), ringwoodite (rw) and pyroxene.  
        The lower mantle (LM) is composed of bridgmaniteand magnesiowustite (mw), and the lower-most mantle (LMM) includes post-perovskite and mw.  
        The proportions of these minerals conserve the Fe/Mg ratio in all mantle layers.
        The core is composed of iron-nickel and an alloy that includes Si. 
        Our model has the flexibility to account for different mineral proportions, different iron contents in the mantle, and different alloys in the core.
        }
        \label{fig:cross-section}
    \end{figure}

    Starting from the center of the planet we consider:
    \begin{enumerate}[I.]
        \item The core may be divided into two layers: outer core and inner core with the transition following the melting line of iron alloy. 
        The outer core may be molten and the inner core solid, similar to Earth (following \citealt{Valencia2006,Valencia2007}). 
        However, for the purposes of this study we have considered the core to be a single layer of iron alloy composed of Nickel $10\%$ by mol and a light element that ranges between 2 and $12\%$ by mol.
        Based on the Earth \citep{Hirose2013} the candidates for the light alloy are O, C, S, H, and Si and they make up at most $15\%$ by mol for the Earth \citep{McDonough1995}.  
        In this study, we focus on Si as the main alloy, given that any presence of it in the core would change the bulk Fe/Si ratio of the planet.   
        We consider a range of $0-10 \%$ Si ($\mathrm{x_{Si}}$) by mol and $2\%$ of a light unspecified alloy: 
        \begin{equation*}
            \mathrm{Fe_{1-x_{Si}}Ni_{0.1}Si_{x_{Si}}} \ .
        \end{equation*}
        Ignoring the inner-outer core transition on Earth, and making the inner core the same composition as the outer core would underestimate Earth's mass by only one part in 10,000 ($\Delta M / M \sim 10^{-4}$).  
        This small discrepancy is due to the small mass of the inner core.  
        Thus, our assumption of a single composition for the whole core for rocky super-Earths is well within the uncertainties of masses and radii.

        \item The mantle is divided into four sub-layers that follow mineral phase boundaries and are determined by the pressure-temperature profile: 

        \begin{enumerate}[A.]
            \item Upper mantle: composed of olivine and pyroxene, in variable proportions set by $\mathrm{x_{py}}$ and variable iron proportions described by  $\mathrm{x_{Fe}}$:
            
            \begin{equation*}
                \begin{split}
                    \mathrm{ (1-x_{py})\left(Fe_{x_{Fe}},Mg_{(1-x_{Fe})}\right)_{2} } & \mathrm{SiO_4}  + \ \\ 
                    & \mathrm{ x_{py}\left(Fe_{x_{Fe}},Mg_{(1-x_{Fe})}\right)_2Si_2O_6 } 
                \end{split}
            \end{equation*}    
                
            \item  Transition zone: composed of wadsleyite (wd) and pyroxene, as well as ringwoodite (rw) and pyroxene. 
            That is, instead of having two layers for the transition between wadsleyite to ringwoodite (higher pressure forms of olivine), as is the case for Earth, we consider one single layer with the mixture of both with equal amounts, justified by the small differences in pressure at which these phase transitions occur.
        
            \begin{equation*}
                \mathrm{(1-x_{py})(rw + wd) + x_{py}\left(Fe_{2x_{Fe}},Mg_{2(1-x_{Fe})}\right)Si_2O_6 }
            \end{equation*}
    
            \item Lower mantle: composed of bridgmanite (bm) and magnesiowustite (mw), in variable proportion according to $\mathrm{x_{bm}}$:
            
            \begin{equation*}
                \mathrm{x_{bm}\left(Fe_{x_{Fe}},Mg_{(1-x_{Fe})}\right)SiO_3 + (1-x_{bm})\left(Fe_{x_{Fe}},Mg_{(1-x_{Fe})}\right)O}
            \end{equation*}
            
            \item Lower most mantle: composed of post-perovskite and magnesiowustite, where post-perovskite is the higher pressure form of bridgmanite and so the same proportions are used as in lower mantle, $\mathrm{x_{bm}}$.
                
        \end{enumerate}
    \end{enumerate}
    
    Recent experimental work by \citet{Niu2015} has suggested the existence of higher-pressure forms for post-perovskite (i.e. $\mathrm{MgSi_2O_5}$) for pressures beyond 1 TPa. 
    Most rocky super-Earths in our sample have mantle pressures below 1 TPa with the exception of the very massive planets that intersect the RTR. 
    For example, Kepler-20b has a core-mantle boundary pressure of 1.8 TPa.  
    Thus, our calculations for these planets are conservative estimates, that may be refined once the equations of states for the high-pressure forms of ppv are known.  
    
    We impose the mantle layers to have the same Mg/Si ratio across all of them, which translates to the condition $\mathrm{x_{py}}=2\mathrm{x_{bm}}-1$ being satisfied for all the compositions we considered. 
    \autoref{fig:cross-section} is a representation of the interior structure of rocky super-Earths employed in our model.  
    
    The total radius $R$ of a rocky super-Earth is dependent on the mass $M$ of the planet and its composition $\chi$, or explicitly in our model, the core mass fraction, and mineral composition represented by $\mathrm{x_{Fe}}$, $\mathrm{x_{py}}$,  $\mathrm{x_{Si}}$:
    \begin{equation}
        R=R(M; \chi) = R(M; \mathrm{cmf}, \mathrm{x_{Fe}},\mathrm{x_{py}}, \mathrm{x_{Si}}).
    \end{equation}
    Notice that a particular combination of values for cmf, $\mathrm{x_{Fe}}$, $\mathrm{x_{py}}$ and $\mathrm{x_{Si}}$ will yield specific values for the Fe/Mg and Fe/Si ratios. 

    Our model differs from other interior structure models \citep{Sotin2007,Grasset2009,Dorn2015,Unterborn2016,Brugger2017} that set the bulk composition first and then solve for the interior structure after.
    For example, \citet{Grasset2009} and \citet{Brugger2017} set a particular Fe/Mg ratio that the planet has to comply with and adjust the different layer thicknesses to fulfill this condition.
    \citet{Dorn2015} use a geochemical model where they choose a bulk composition a-priori (e.g. pyrolite), and then solve for which minerals are present and in what proportions according to a Gibbs free energy minimization treatment.  
    In comparison, our model can get to a desired composition via an inversion scheme.  
    By not considering any of the Al and Ca bearing pyroxenes or perovskites, our model obtains the largest Mg content, or minimum cmf. 
   
    We solve the differential equations for density, pressure, gravity, mass and temperature following \citealt{Valencia2006,Valencia2007}, and use the Vinet equation of state (EOS,  \citet{Vinet1989}) with a thermal Debye correction.
    With six parameters describing the behaviour of the material at reference pressure ($P_0=0$) and temperature ($T_0=300$K), namely: the density $\rho_0$, the bulk modulus $K_0$, the first derivative of the bulk modulus $K'_0$, the first Gruneisen parameter $\gamma_0$, the second Gruneisen parameter $q$ and the Debye temperature $\theta_0$.  
    The temperature structure is self consistently obtained with a parameterised convection treatment and benchmarked to the Earth's potential mantle temperature.
    It follows a conductive profile within the top and bottom boundary layers of the mantle, and the adiabatic gradient in the bulk of the mantle and throughout the core (see \citet{Valencia2006} for details). 
    {This treatment assumes mobile lid convection. Below we quantify the effects of temperature on the structure, and show that the effects are small ($\Delta R < 1\%$), suggesting other parametrizations (e.g. stagnant lid) are equally valid.}
    
    To obtain the relevant equation of state for the mineral mixture at each of the layers we use a linear volume mixing model \citep{Badro2007}, with density calculated as    
    \begin{equation}
        \rho_{\mathrm{mix}} = \frac{\sum_i x_i \mu_i}{\sum_i (x_i \mu_i)/\rho_i},
        \label{eq:rho_mis}
    \end{equation}
    where $\rho_i$ are the density of each of the components to mix,  $x_i$ the atomic percents and $\mu_i$ the atomic mass units.
    
    {With the parameters for each end-member, we create a grid of density as a function of pressure and temperature $\rho_{\mathrm{mix}}(P, T)$ using \autoref{eq:rho_mis}.  
    For example, for the lower mantle, we mix four components: the Mg and Fe end-members for bridgmanite and magnesiowustite, in appropriate proportions.}

    With the mixture data $\rho_\mathrm{mix}(P,T)$, we then fit six EOS parameters -- $\rho^{\mathrm{mix}}_0,$ $K^{\mathrm{mix}}_0,$ $K'^{\mathrm{mix}}_0, \gamma^{\mathrm{mix}}_0,$ $q^{\mathrm{mix}}, \theta^{\mathrm{mix}}_0$ .
    In the case of the core, following \citet{Morrison2018} the last three parameters are fixed for the iron-nickel end-member. 
    
    The most unconstrained parameters were the three thermodynamic ones given that thermal effects have a smaller contribution to density compared to pressure effects.   
    The parameters that had the largest co-variance (negative) where $K^{\mathrm{mix}}_0$ and $K'^{\mathrm{mix}}_0$.
    It is worth noting that the end-members we considered in the core are $\mathrm{Fe_{0.9}Ni_{0.1}}$ and $\mathrm{Fe_{0.8}Ni_{0.1}Si_{0.1}}$ following \citet{Morrison2018}. 
    Thus, to avoid issues with extrapolation and in line with the proportion of light alloy the Earth's core is thought to have, we restricted the possible variation of Si in the core up to $0.1$ by mol.

\begin{figure}
    	\includegraphics[width=\linewidth]{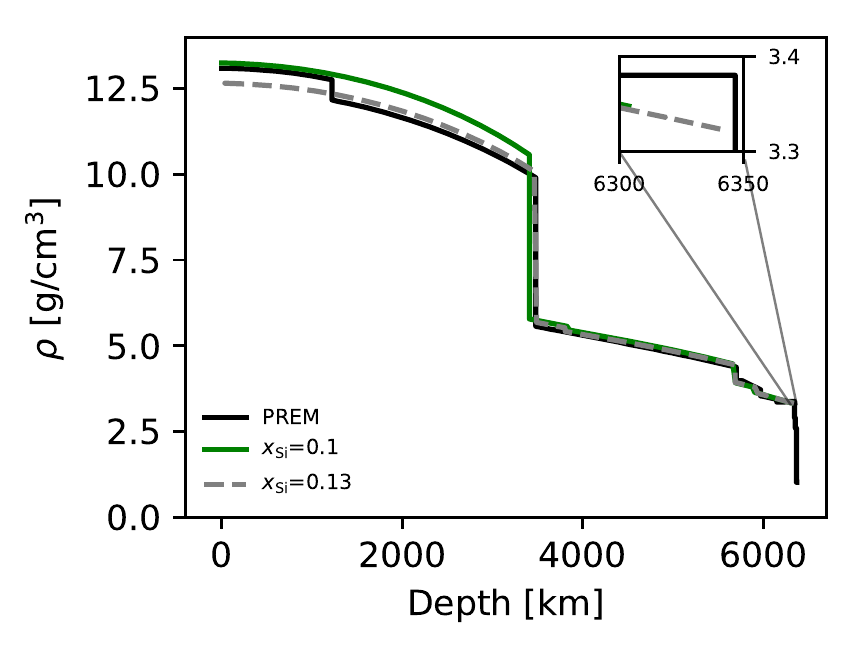}
        \caption{Density profile for a $1\,M_\oplus$-planet with 0.1 (green) and 0.13 (dashed grey) by mol of Si in the core, and cmf=0.325. 
        Earth's internal structure from PREM is shown for comparison.}
        \label{fig:P-chem}
    \end{figure}

    We tested our model against the Preliminary Reference Earth Model (PREM, \citet{Dziewonski1981}) for Earth. 
    The results are shown in \autoref{fig:P-chem}. 
    To better fit the Earth with our limited chemical inventory, we set the amount of iron in the mantle to be 0.07 instead of the nominal $\sim 0.1$ (given the $\mathrm{Mg}\#=89$) thought to apply to Earth \citep{McDonough1995}. 
    \autoref{Table:EOS_param} shows the EOS parameters for each of the layers assuming this particular composition.
 
    \begin{table}
        \begin{tabular}{lcccccccr} 
        \hline
    	Compound&  $\rho_0$ & $K_0$ & $K'_0$ & $\gamma_0$ & $q$ & $\theta_0$ \\
    	
    	\textit{Mantle$^\dagger$}\\
    	\hline
        UM & 3323.5 & 126.4 & 5.47 & 1.12 & 0.2 & 619 \\
        TZ & 3512.5 & 157.8 & 4.64 & 1.27 & 1.5 & 665 \\
        LM & 4105.8 & 214.7 & 4.85 & 1.89 & 1.3 & 698 \\
        LLM & 4111.0 & 202.8 & 4.52 & 1.89 & 1.3 & 454 \\
        \textit{Core$^\ddagger$}\\
        \hline
        $\mathrm{Fe_{0.9}Ni_{0.1}}$ &  8278.9 &  157.5 &      5.61 &     2.0 &  1.0 &  417  \\ 
        $\mathrm{Fe_{0.8}Ni_{0.1}Si_{0.1}}$ &  7720.1 &  125.2 &      6.38 &     2.0 &  1.0 &  417  \\
        \hline
    	\end{tabular}
        \centering
    	\caption{{Parameters of the resulting EOS of the mixture at each structure layer corresponding to a specific composition of 0.07 by mol of iron, 0.5 of pyroxene and 0.25 of magnesiowustite. Each composition considered in the MCMC had different EOS values.} \protect\\
    	$\dagger$ -- Mantle values are from \citet{Stixrude2011} work.
    	\protect\\ 
    	$\ddagger$ -- Core values are from \citet{Morrison2018} work.}
    	\label{Table:EOS_param}
    \end{table}

\subsection{Effect of Differentiation,  Temperature and Fe Partition on Radius and Mass}
    During planet formation, rocky planets are thought to differentiate into an iron core in the centre, overlain by a magnesium-silicate mantle, owing to the fact that iron is heavier.  
    As long as the planet has a partially molten interior, the iron can flow to the core differentiating the planet and carrying with it siderophile elements \citep{Fiquet2010}.  
    The halfnium-tungsten radioactive clock suggests that $90\%$ of the Earth's core had formed within 30 \unit{Myr} \citep{Jacobsen2005}. 
    However, some iron may be left in the mantle if the timescale to cool below the melting point of iron is faster than the sinking timescale, and thus, can be different for some planets.

       Given that we do not know how differentiated super-Earths may be, we quantify the effects of differentiation in the total radius of planets.
    \citet{Valencia2009} had shown that the difference between a differentiated earth-like composition and an undifferentiated one where the bulk Fe/Si fraction is preserved is $2\%$ in radius for a 10$M_\oplus$ planet.
    In this study, with improved equations of state, we quantify this effect by changing the amount of cmf and iron in the mantle, as well as the amount of pyroxene to olivine (and hence, mw to bm/ppv) to keep both Fe/Si and Fe/Mg the same despite different degrees of differentiation.
    
    \begin{figure}
        \includegraphics[width=\linewidth]{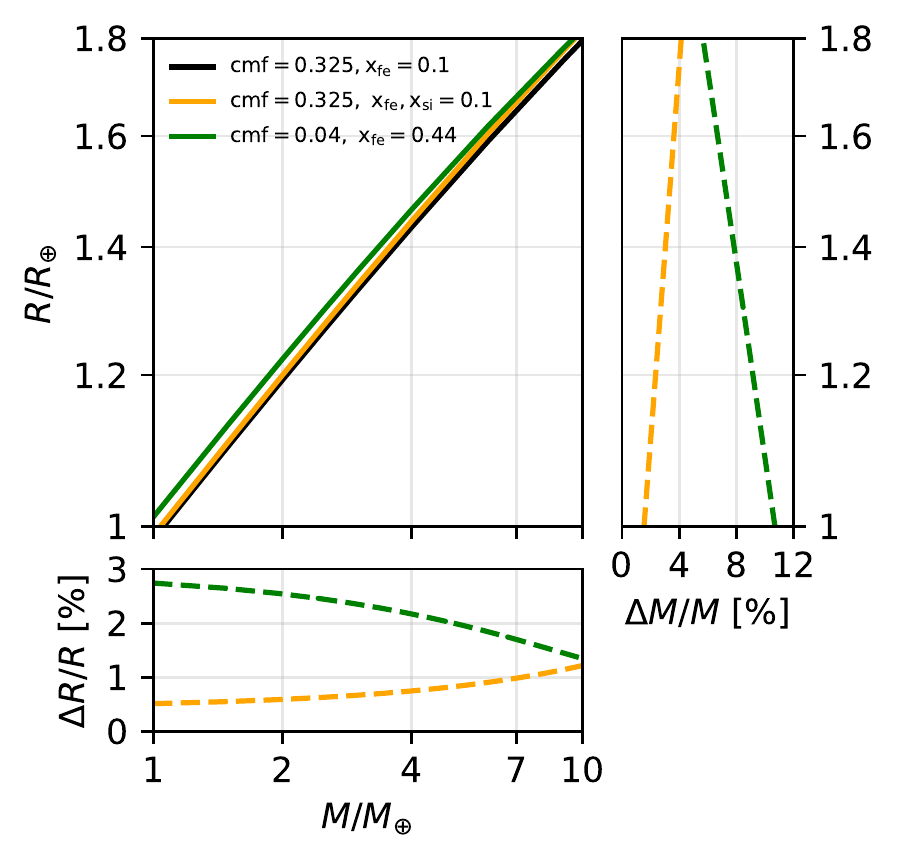}
        \caption{Effects of differentiation on the mass-radius relationship of rocky planets.  
        Black:  cmf of 0.325 and iron in the mantle $x_\mathrm{Fe}= 0.1$ by mole fraction, and no Si in the core ($\mathrm{x}_\mathrm{Si}= 0$) corresponds to Fe/Si = 2 and Fe/Mg = 2.  
        Orange: cmf = 0.325,  $\mathrm{x}_\mathrm{Fe}= 0.1$,  $\mathrm{x}_\mathrm{Si}=0.1$ and Fe/Si = 1.7 and Fe/Mg = 1.9.
        Green: cmf = 0.04,  $\mathrm{x}_\mathrm{Fe}= 0.44$, $\mathrm{x}_\mathrm{Si}= 0$ and Fe/Si = 2 , and Mg/Si = 2.}
        \label{fig:FeMan}
    \end{figure} 
    
    We show the results in \autoref{fig:FeMan}.
    The difference in radius due to differentiation is $\Delta R \sim $  3, 2 and $1.4 \%$ for a 1, 5 and 10 $M_\oplus$, respectively.
    Conversely, the difference in mass due to differentiation (if radius is kept constant) is $\Delta M/M \sim$ 11, 8 and $6 \%$ for a 1, 1.4 and 1.8 Earth-radii planets. 
    The errors in radius and especially mass in observations are yet too large to discern the effect of differentiation.

    We also calculate the effect of surface temperature, on the planet's radius.
    We use both 300K and 1000K surface temperature {with corresponding potential mantle temperatures of 1600K and 2200K, respectively}, over a range of possible masses and cmfs. 
    As expected, the differences are even smaller than differentiation ($\Delta R/R \sim 0.5\%$), owing to the small values of thermal expansion of rocks.  
    {Meaning that rocky planets that are highly irradiated and partially molten and those that are poorly irradiated at long distances have similar radii. 
    This stands in contrast to planets that have a volatile envelope.}
    
    {Finally, we look at the effects of iron partitioning in the mantle between the most dominant minerals.
    Experiments by \citet{Auzende2008} suggest that bridgemanite and post-perovskite may have a different iron content compared to magnesiowustite.  
    To quantify this effect on the radius of an Earth-like planet (cmf=0.325), we changed the amount of iron between these two mineral phases while keeping the ratio of 1:4 magnesiowustite to bridgmanite(or post-perovskite), as well as the bulk iron content, the same. 
    Thus, to compare to our fiducial composition with iron number $\mathrm{x_{Fe}}=0.1$ throughout, the compositions considered had to satisfy the relation $0.8 \mathrm{x_{Fe}^{bm}} + 0.2 \mathrm{x_{Fe}^{mw}}=0.1$.
    
    Under these conditions, the maximum effect of iron partitioning ($\Delta R/R \le 0.5\%$) was obtained with the extreme composition of no iron in bridgemanite/post-perovskite (MgSiO$_3$) --  corresponding to $\mathrm{x_{Fe}^{mw}}=0.5$.  
    However, for Earth, values of $\mathrm{x_{Fe}^{mw}}$ fall between 0.2-0.35 \citep{Muir2016}. Thus, it is possible the effect on radius to be lower than $0.5\%$. 
    We conclude that the influence of Fe partitioning on planetary radius is not significant for our purposes ($\Delta R/R \lesssim 1\%$), and assume the same iron number throughout all minerals, $\mathrm{x_{Fe}}$.}

    {The implications are that the radius of a rocky planet is highly sensitive to the amount of total iron (both in the core and mantle) and Mg-Si rock material, and much less to other parameters such as temperature, differentiation degree (amount of iron in the mantle vs. core) and iron partitioning in the mantle.}

\subsection{Calculating Composition: MCMC}
    The interior structure code allows us to pose a forward problem:  given a mass and composition, we calculate the radius.
    However, our interest is to calculate the composition given a mass and radius with associated uncertainties.  
    To this end, we use the affine invariant Markov Chain Monte Carlo (MCMC) sampler \texttt{EMCEE} \citep{Foreman2013} and couple it to our interior structure code.  
  
    We choose a log-likelihood function, $\log(p)$, that depends on the observed $M_\mathrm{obs}$ and $R_\mathrm{obs}$, as well as their uncertainties $\sigma_M$ and $\sigma_R$, respectively:

    \begin{equation}
        \log(p) =  -\frac{\left(M_\mathrm{obs} - M \right)^2}{2\sigma_{M}^2} - \frac{\left(R_\mathrm{obs} - R \right)^2}{2\sigma_{R}^2}.
    \end{equation}
    
    The planetary radius is calculated with the interior structure model, $R(M,\chi)$.
    We sample the planetary mass $M$ and composition $\chi$(CMF, $\mathrm{x_{Si}}$, $\mathrm{x_{Fe}}$) to maximize the log-likelihood function and obtain the probability density distributions of these parameters for each planet.   We recognize that both planetary radius and mass may not be entirely independent and leave more sophisticated statistical analysis for future work.
  
    We look at different levels of complexity in the compositions considered.
    For case I we only vary cmf between 0 and 1, while keeping all other parameters constant and set to Earth-like values while having no silicate in the core ($x_{\mathrm{Fe}}=0.1$, $x_{\mathrm{py}}=0.6$, $\mathrm{x_{Si}}=0$).  
    This case is similar to the one considered by \citet{Hardcore} for a hypothetical planet. 
    
    In case II we also allow for Si to be present in the core in the range of 0-0.1 by mol, while also varying the cmf, but fixing the amount of iron in the mantle.  
    Case III consisted in varying cmf, $\mathrm{x_{Si}}$ as well the amount of iron in the mantle ($\mathrm{x_{Fe}}$) between 0-0.2. 
    The last case amounts to considering also different degrees of differentiation. 
   
    For one particular planet as a test, Kepler-10b, we also considered a fourth case where we varied $\mathrm{x_{py}}$ between 0 and 1.  
    The results of including this case yielded different estimates in bulk Fe/Si and Fe/Mg of only 0.04 absolute differences in the mean values.  
    Changing the amount of $\mathrm{x_{py}}$ amounts to considering different Si/Mg values for the mantle. 
    \citet{Delgado2010} proposed that planets around Si/Mg-rich stars would have pyroxene-rich mantles. 
    Unfortunately, the EOS of pyroxenes and olivines is too similar for the different proportions to affect the radius of the planet enough to be seen in the data ($\Delta R/R \sim 0.02\%$). 
    
    For each exoplanet we considered, we assume uniform priors for all of these parameters.
    Furthermore, the planetary mass and radius are initialized following Gaussian distributions described by the observed data, whereas the chemical parameters $\mathrm{x_{Si}}$, $\mathrm{x_{Fe}}$ and $\mathrm{x_{py}}$ follow a uniform distribution. 
    
    Due to computational limitation, we evolve 64 random walkers with $\sim 50$ burn in steps. 
    We generate the posterior from the next 400 steps and calculate the geometric mean and the maximum a-posterior probability (MAP or mode) values with the corresponding $1\sigma$ confidence intervals.  
    
    \begin{figure*}
    \centering
        \includegraphics[width=\linewidth,height=0.9\linewidth]{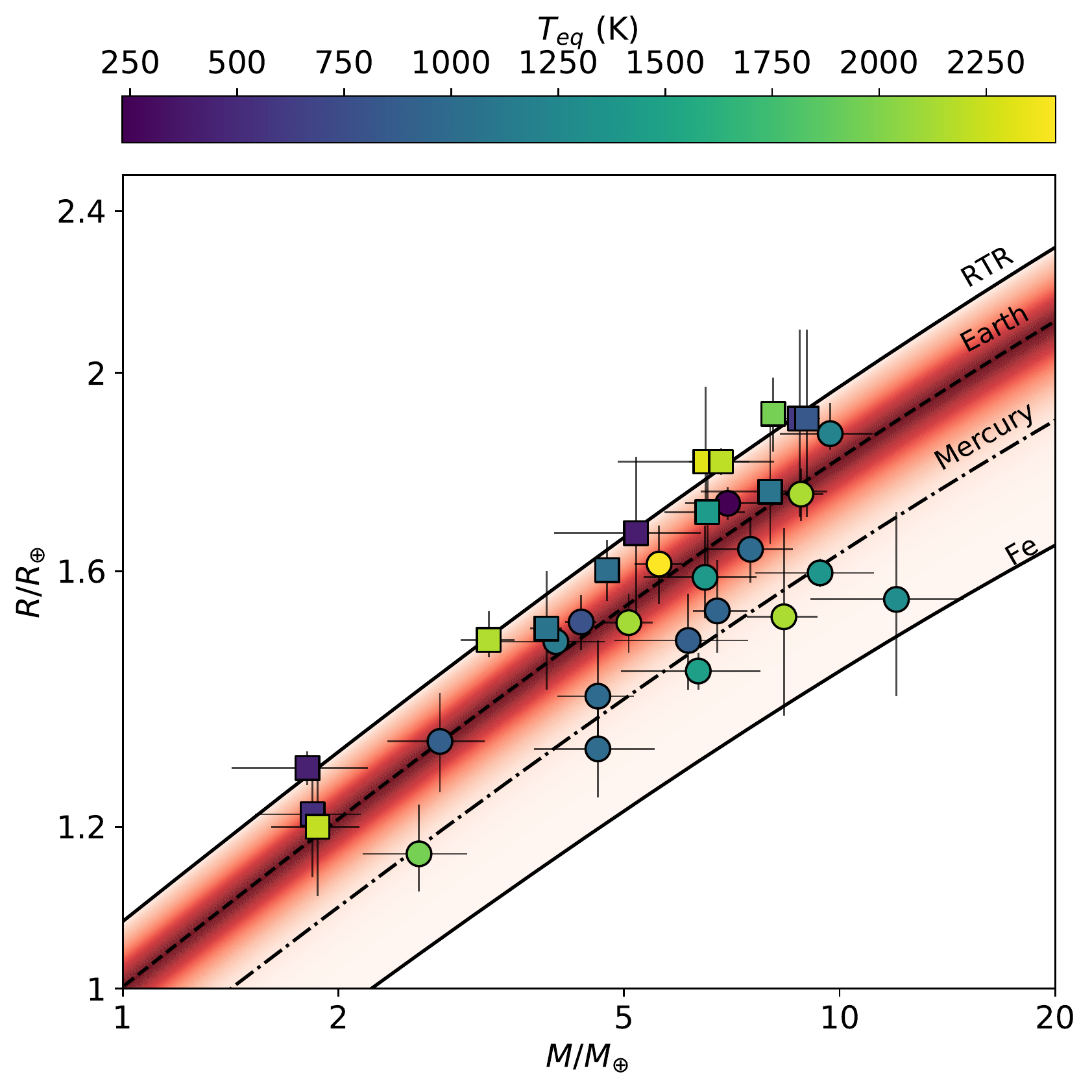}
        \caption{Exoplanets sample used in the study ($\Delta M, \Delta R \leq 25 \% $). 
        Exoplanets are colour coded according to their equilibrium temperatures.
        Dark lines show the mass-radius relationships for specific compositions including pure iron-nickel, planets similar to Mercury (cmf=0.63), Earth (cmf=0.325) and the Rocky Threshold Radius (RTR) which is the largest size a rocky planet can have (with no iron). 
        Red shaded region are the compositions of stars obtained from their Fe/Si and Fe/Mg ratios and translated into planet's composition.
        Circles: exoplanets with $1\sigma$ confidence interval intersecting the RTR, squares: exoplanets with $1\sigma$ confidence interval within the rocky region and not intersecting the RTR.
        The stars composition is confined to a smaller mass-radius region than where exoplanets have been found.
        }
        \label{fig:RvsM}
    \end{figure*}
    
\section{Results}
\subsection{Rocky Exoplanet Sample}
    For meaningful results, we restrict our sample to planets that are in the rocky region, have both measured mass and radius and the uncertainties in mass and radius are below 25\% each. 
   
    Using the NASA exoplanet archive\footnote{ \url{https://exoplanetarchive.ipac.caltech.edu} } we obtain 33 planets that fit these criteria, with 12 having more than one measurement.
    For those having multiple observations, we generally choose the most recent data from the archive.
 
    For K2-106b (also named EPIC~220674823b)  and Kepler-105c planets, we use estimates made by \citet{ref_EPICb_Guen} and \citet{ref_Kepler-105c} respectively, since their reported radii uncertainty is lower compared to other groups.
    We choose not to include planets K2-38b, Kepler-78b, Kepler-93b from \citet{Xie2014} and \citet{Stassun2017} due to the fact that the masses and their uncertainties seem to be overestimated and differ significantly compared to other studies.

    These planets are shown in \autoref{fig:RvsM}, as well as the mass-radius relationships for silicate rock (no iron) which sets the Rocky Threshold Radius (RTR), the Earth (cmf = 0.325 \citet{Wang2018}), Mercury (calculated by \citet{Hauck2013} to be cmf = $0.63\pm 0.03$  for an Fe+Si core) and a pure Fe-Ni planet.  
    Notice that the Rocky Threshold Radius is calculated using all appropriate mineral phases. 
    This is a more accurate treatment compared to assuming only MgSiO$_3$, commonly used in other works, that leads to an underestimation of 80 \unit{km} for 1 M$_\oplus$ planet with no core.
    We have ignored all planets that lie completely above the RTR, as this requires the planet to have volatiles in large enough quantities to affect its radius, and hence, its composition cannot be uniquely determined with only mass and radius measurements.  
    We have also denoted the planets that can intersect the RTR by examining the $1\sigma$ confidence interval on mass and radius observations. 
    These planets have the highest probability of being volatile within our sample.
    Again, given that we cannot rule out the possibility of an envelope with mass-radius pair measurements, our estimates of refractory ratios of Fe/Mg, Fe/Si and cmf correspond to minimum values. 
    
    We have shown the results of our MCMC simulation combined with interior modeling for 55 Cnc e in \autoref{fig:corner} as an example of our procedure employed with the 33 exoplanets in our sample. 
    This particular case was obtained varying the amount of Si and Fe in the core and mantle, respectively (compositional case III).
    
    We show the MAP and the 1$\sigma$ values obtained for mass, radius, cmf, amount of Si in the core, and amount of Fe in the mantle, and derived chemical ratios Fe/Si and Fe/Mg. 
    As expected, the most well-constrained compositional parameter is the cmf owing to the fact that for a given mass it influences the radius the most, compared to the other two parameters, $\mathrm{x_{Si}}$ and $\mathrm{x_{Fe}}$, within the bounds considered. 
    This can be seen as well from the strong correlation between radius and cmf in the corner plot (\autoref{fig:corner}). 
    
    We employ this procedure for all planets and summarize the results in  \autoref{Table:summary1} and \ref{Table:summary2}, which correspond to planets that are well within the rocky region, and those that intersect the RTR and may be volatile, respectively. 
    We find that the difference between the different compositional cases is small, and therefore, we only show the results for the most general compositional case III.  We also show the marginal distribution of the Fe/Si ratio of each planet in \autoref{fig:FeSi_sample}.
    The observational radius and mass reported in \autoref{Table:summary1} and \ref{Table:summary2} may differ from the data obtained by our MCMC treatment given that we assume Gaussian errors for mass and radius, but more importantly, because we assume planets are rocky. 
    This means we limit our posterior to exclude values that lead to planets above the RTR and below the Fe-Ni composition.

    \begin{figure}
    	\includegraphics[width=\linewidth]{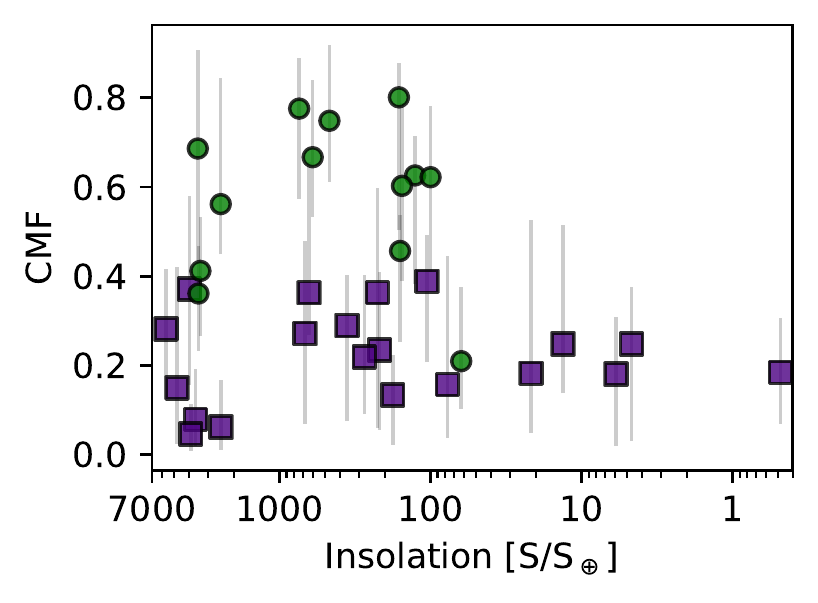}
        \caption{Core-mass fraction as a function of received stellar flux in term of Earth's received flux. 
        Purple squares: planets that are in the rocky region and intersect the RTR; Green circles: planets that are in the rocky region and do not intersect the RTR (within $1\sigma$). 
        Given the observational bias against long period compact planets, there seems to be no trend between core-mass fraction and received flux.
        A few RTR-crossing planets have very high insolation values and are consistent with small cmf, posing a challenge to formation theories.
        }
        \label{fig:cmf_P}
    \end{figure}  
    
    To investigate possible trends in composition with solar insolation, we show the results of cmf for all the planets as a function of star flux received, \autoref{fig:cmf_P}. 
    If we assume all the planets in our sample are rocky, we find that planets with high cmf are absent at low insolation values. 
    However, this may be due to observational biases from the Kepler data set at large periods.
    We make a point of distinguishing by colour and symbol the planets that may be volatile and intersect the RTR (i.e. RTR-crossing planets) from those that are more likely to be rocky and do not intersect the RTR (i.e RTR-embedded planets).
   
    There is an intriguing handful of RTR-crossing planets at very high insolation values that are iron-deficient with respect to Earth: 55 Cnc e, HD-80653b, K2-131b, WASP-47e.  
    Should they be volatile instead of being rocky (with a higher cmf for their rocky interior), the composition of their envelope would be key to understanding formation models.
    If their atmospheres are H/He dominated, one would have to explain how they avoided atmospheric evaporation. 
    On the other hand, if they are water dominated, one would have to explain their origin given how close they are to their parent star. 
    
    In contrast, LHS 1140b is a planet within the rocky region (or RTR-embedded) with a minimum cmf of $\sim 0.2$ and lower insolation compared to Earth ($\sim 0.5S_\oplus$).
    With the low received stellar flux, this planet may have a liquid/solid water layer, and if so, the cmf would actually be larger. 
        
    As more data arrives, especially with the TESS mission and the different instruments coming online that are capable of measuring the masses of small planets around M Dwarfs we will continue to build this data set and investigate any trends in composition for super-Earths.
   
    \begin{figure*}
    \centering
    	\includegraphics[width=\linewidth,height=0.8\linewidth]{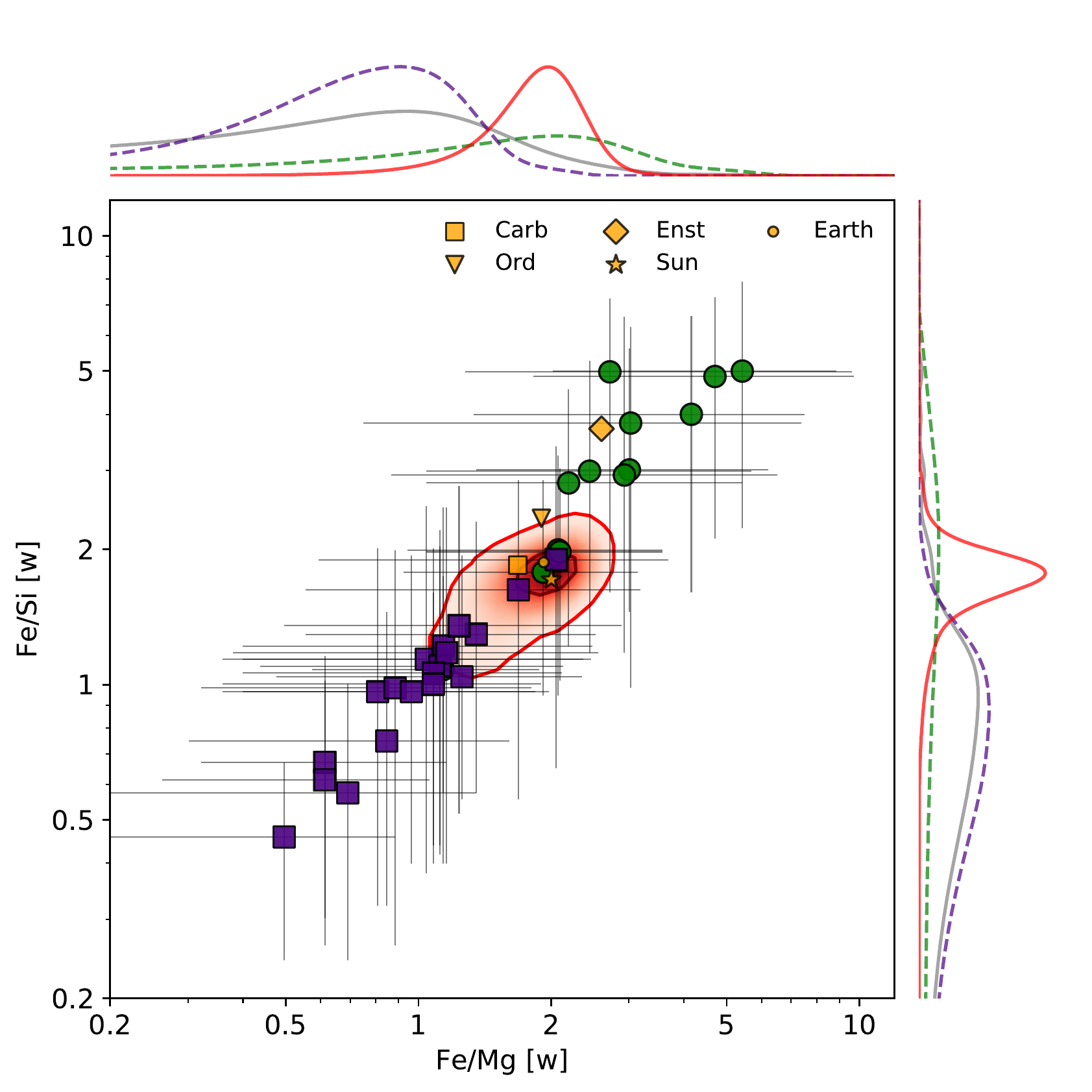}
        \caption{Star-planet refractory composition comparison.  
        Chemical ratios of planet-hosting stars are shown in red at contours of 1 and 2 $\sigma$ values, compared to all the exoplanets grouped into RTR-crossing (purple squares) and RTR-embedded planets (green circles). 
        Notice the axes are in logarithmic scale. The marginal distribution of Fe/Mg and Fe/Si are shown in the perimeter of the scatter plot for stars (red), all exoplanets (grey) and exoplanet sub-samples. 
        Solar system objects are shown in yellow. 
        Planets' refractory ratios span a wider range than stars.
        }
        \label{fig:Star_Pl}
    \end{figure*}  

\subsection{Star-Planet Population Comparisons}
    All planets in our selected sample, with the exception of four, have host stars with undetermined chemical composition.  
    Therefore, we perform chemical comparisons between planets in our sample and that of stars as a population by using the stellar abundances of planet-hosting stars from the Hypatia Catalogue \citep{Hinkel2014}.
    The stellar data from Hypatia usually has more than one reported measurement of a chemical abundance for a given star, due to multiple observational reports.
    We use the mean values for the chemical abundances for a given star and obtain the associated error by assuming the calculated abundance follows a Gaussian distribution.
    From this, we obtain the stellar chemical ratios Fe/Mg and Fe/Si.
    \autoref{fig:Star_Pl} shows the absolute ratios Fe/Mg and Fe/Si by weight (not normalized to the Sun) of the stars (in red), and those of the planets in our sample.
    We also show the values for the Sun's photosphere, the Earth's composition, and of a few different chondrites that show the variation in composition in the Solar System. 
    Note that the figure is in logarithmic scale. 
    The stars span a much wider range in refractory ratios than samples in the Solar System, except for enstatite chondrites that are known for being iron-rich. 
    Fitting Gaussian distribution to the Fe/Si (Fe/Mg) stellar chemical ratio yields means of 1.69 (1.78) and variances of 0.11 (0.13).

    We show as well the results for each of the planets in our sample in \autoref{fig:Star_Pl}, and distinguish between the RTR-crossing and the RTR-embedded planets. 
    The planets align with a constant Mg/Si line (of 1.04 by weight, similar to the Earth) by construction.
    The spread around this value comes from allowing additional Si to be in the core, while preserving the Mg/Si ratio in the mantle. 
    In comparison, the stars have a mean value of Mg/Si of 0.95 by weight.  
    Notice that in this space of chemical ratios the ranges can span from 0 to infinity accentuating differences in a non-linear way.  

    From \autoref{fig:Star_Pl}, it is clear that planets span a wider range of chemical ratios than those of the stars. 
    Although, error bars are large, especially for the iron-rich planets.

    To perform a more quantitative comparison, we obtained the probability density distribution of the planets, by performing a weighted kernel density estimation (KDE) on the trimmed mean.   
    We chose to trim the $10\%$  upper quantiles of each of the planets' inferred compositional distribution, to arrive at the trimmed mean for each planet.  
    The motivation is that although the internal structure code allows for pure iron-planets, these extreme compositions are unlikely, and correspond to infinite chemical ratios,  which can artificially skew the distribution.  
    Without trimming the data, the means are $15\%$ smaller,  the medians a few per cent  smaller, and the modes are unchanged, compared to using the trimmed distributions. 

    The bandwidth for the individual kernel was chosen by implementing a cross-validation approach with an out-of-training sample of 4. 
    Subsequently, the kernels were weighted by the inverse of their error before being added. 
    The KDEs for the trimmed means for Fe/Si and Fe/Mg are shown in the perimeter of \autoref{fig:Star_Pl} for all the planets, as well as the RTR-crossing and RTR-embedded planets separately, in addition to the distribution of stars.
    The distribution of planets is clearly much wider than those of stars.
    Notice, that the logarithmic scale accentuates differences at low Fe/Si and Fe/Mg values, and reduces differences at high values. 
    Notably, the distribution of planets includes higher Fe/Si and Fe/Mg values than for stars.

\begin{figure}
    	\includegraphics[width=\linewidth]{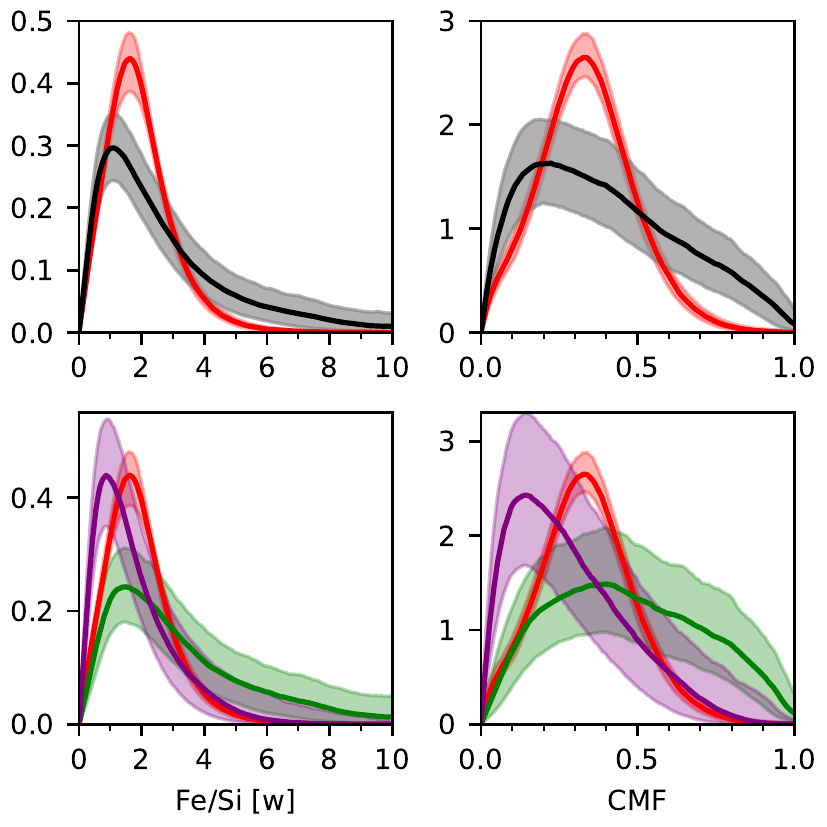}
        \caption{The probability density distributions for core mass fractions (right) and Fe/Si ratios (left) for stars (red) and exoplanets obtained by re-sampling kernel density estimates. 
        The first row shows the exoplanet population as a whole (grey), and the second row shows the KDEs grouped into RTR-crossing (purple) and RTR-embedded (green) exoplanets.
        {Comparing the probability densities in cmf space (right panel) showcases the differences between the actual planets and the expected composition from a primordial origin (stars) more clearly}.
        }
        \label{fig:KDE}
    \end{figure}

    Another way to compare the distribution of planets to those of the stars is to bootstrap different KDEs obtained from sampling the distribution of each planet.  
    That is, we sample each of the compositional distributions of the 33 planets (e.g. from \autoref{fig:FeSi_sample}) and with this group obtain a KDE. 
    We repeat this procedure 1000 times to obtain a mean KDE, and $1\sigma$ confidence interval around this distribution.  
    This method has the advantage of taking into account the spread in the compositional distribution of each planet, which arises from the uncertainty in the mass and radius measurements.  
    The results are shown in \autoref{fig:KDE} both in the chemical ratios (Fe/Si and Fe/Mg) and the cmf linear space, while considering the planets as a whole or dividing them into RTR-crossing and RTR-embedded planets.

    If we consider all the planets in our sample, including those that intersect the RTR and may be volatile,  the distribution of planets is much wider than that of stars. 
    The planet's distribution peaks at lower values for Fe/Si and Fe/Mg than for stars, but has a long tail that extends to higher Fe/Mg and Fe/Si values than stars. 
    However, the peak at lower values is being determined by the RTR-crossing planets, which may not be rocky. 
    By decomposing the KDEs between RTR-crossing and RTR-embedded planets, we find that if rocky, the former has a two-fold iron depletion with respect to the stars as seen in their Fe/Si ratio (mean values are 0.86 vs 1.69, for RTR-crossing planets and stars respectively). 
    At the moment, we lack a reliable theory that forms massive iron depleted planets, or super-Moons \citep{Scora2020}. 
    Thus, it may be an indication that these planets are not all rocky.
    
    A better comparison space may be core-mass fractions. 
    For this, we translate the chemical ratios of the stars to the composition of planets with the same chemical ratios denoting a primordial composition.  
    To perform this translation,  we considered compositional case III. 
    By allowing $\mathrm{x}_\mathrm{Fe}$ and $\mathrm{x}_\mathrm{Si}$ to vary, each stellar composition will have a range for cmf.

    By looking at the whole sample of planets, there is clearly a much wider distribution than expected, had they form primordially. 
    The distribution of planets is significant at values above cmf=0.5, whereas the primordial composition drops off steeply beyond this value.  

\begin{figure}
    	\includegraphics[width=\linewidth]{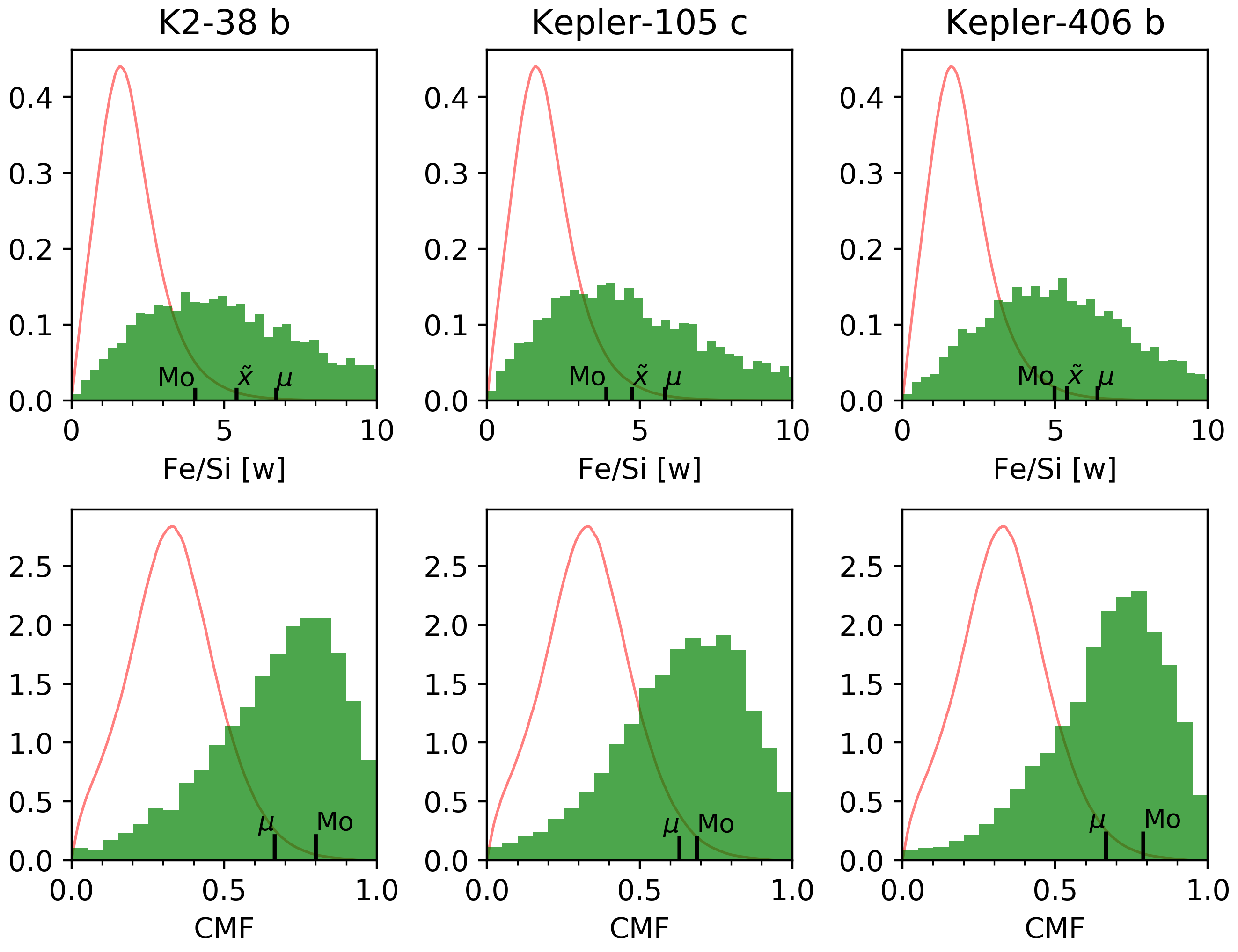}
        \caption{Marginal distribution of Fe/Si and cmf for the most iron rich planets in our sample (cmf$\sim0.8$) compared to the stars' distribution of the means (red).
        $\mu,\mathrm{Mo},\Bar{x}$ indicate the position of mean, MAP (mode) and median of the distribution. 
        In cmf space the mean and median are very similar. 
        These planets may be delineating the maximum iron enrichment attained during planet formation.
        }
        \label{fig:iron_rich}
    \end{figure}

    There are planets with large cmfs (with a mean value of $\sim 0.8$) that are influencing the planets' distribution and warrant special attention, namely K2-38b, Kepler-105c and Kepler-406b.
    Improving the mass estimates of these super-Mercury planets is a fast way to test the chemical planet-star connection. 
    We have shown in \autoref{fig:iron_rich} the distributions of each of these planets compared to the stars population. 
    While the MAP of these planets is more in line with the stars' composition, the median and mean suggest a different origin. 
    But even when considering the MAP values, these planets' compositions correspond to the upper 97 percentile for Fe/Si, and 99.5 percentile for cmf of the stars' distributions, suggesting there may be high planetary processing during formation in exoplanets.  

    Comparatively, Mercury's high iron content within formation scenarios is thought to come from a giant impact \citep{Benz2007}  but details have yet to be explained. 
    Recent calculations by \citet{Clement2019} obtain Mercury's mass, composition and period only in only $1\%$ of their N-body simulations. 
    Consequently, it appears essential to also determine reliably how many super-Mercuries are in the exoplanet sample as to compare solar and extrasolar formation theories.  
    The pathway includes obtaining more precise mass measurements.  
    
    Translating the composition of stars into planets has the added advantage that it can be shown in a mass-radius diagram. 
    We have done so in  \autoref{fig:RvsM}.  
    We used the mean value for the stars chemical ratios and the spread around it from different measurements, and assume compositional case III. 
    Notice that the region occupied by the composition of the stars, which would denote a primordial composition is narrow.    
    This is because iron content has to be substantially modified to have an impact on the structure and radius of a planet. 
    That is, to change the radius of a planet by $10\%$ the Fe/Mg ratio needs to change by a factor of $\sim 6$ for a $1 M_\oplus$-planet. 
    In fact, if we only consider the mean values of the stellar chemical ratios without the error bars arising from different reported stellar compositions, the spread in cmf is much narrower.  
    This exemplifies the need for better chemical constraints on stars as well. 

    It is clear that many planets sit outside the range of stellar compositions, but that their error bars are too large to make definite conclusions.

 \subsection{Uncompressed Densities}
\begin{figure}
    	\includegraphics[width=\linewidth]{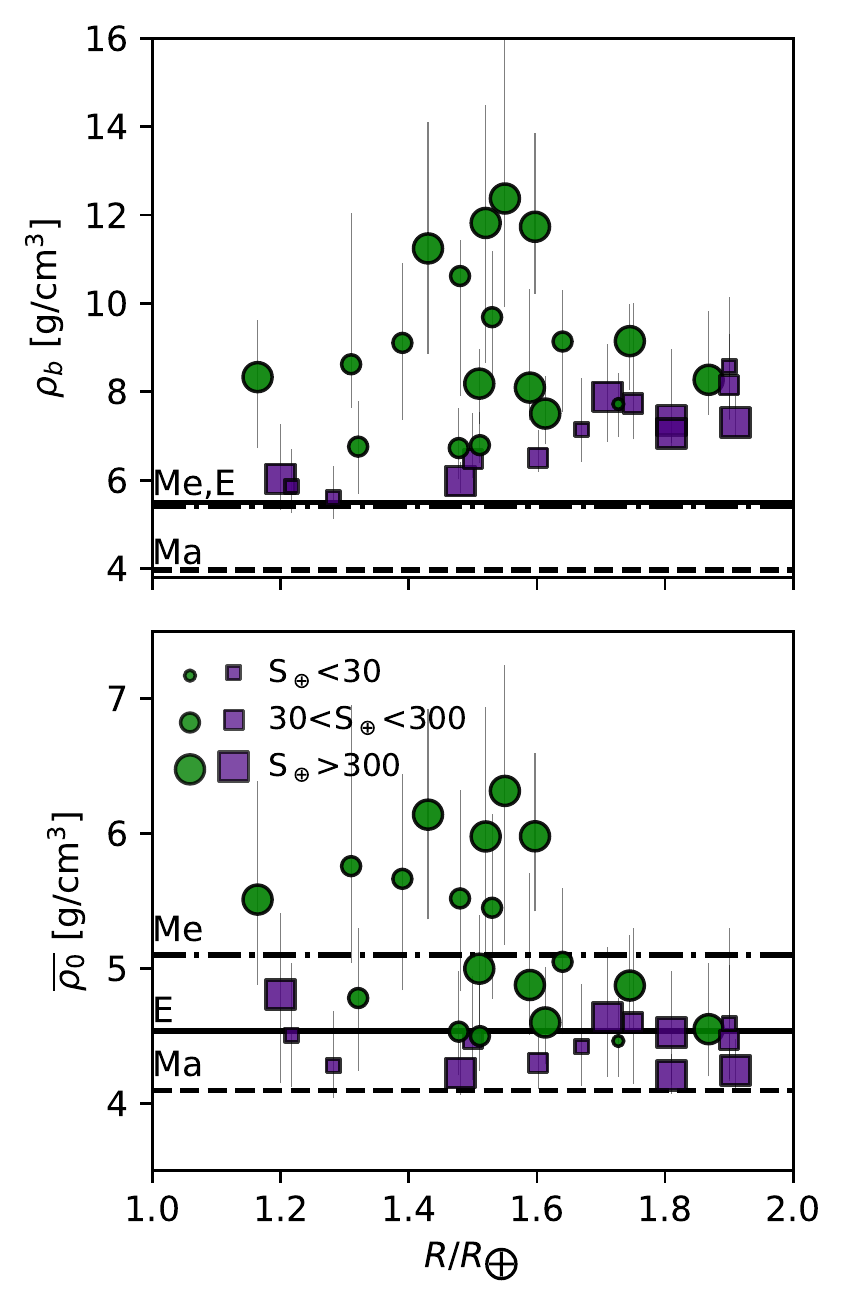}
        \caption{Uncompressed ($\overline{\rho_0}$) and bulk ($\rho_b$) densities of the exoplanet population within the rocky region with masses and radii errors below $25\%$. 
        Differences in bulk density can arise from differences in composition, and pressure regime, whereas differences in uncompressed density originate only from differences in composition. 
        Densities for Mercury (Me), Earth (E) and Mars (Ma) are shown for reference.   
        Size of the symbols represent different received fluxes in terms of the Earth's received flux. 
        Colours and symbols are the same as in \autoref{fig:Star_Pl}.  
        There is an absence of planets at $R>1.65 R_\oplus$ and $\overline{\rho_0}>5$ \unit{g/cm^3}, which may indicate the parameter space where runaway growth takes over during formation. 
        If highly irradiated compact planets are the result of atmospheric evaporation, iron enrichment and perhaps depletion is set before gas dispersal.}
        \label{fig:rhovsR}      
    \end{figure}

    Another useful metric to compare planets is the uncompressed density, $\overline{\rho_0}$. 
    This is the density that a planet would have if all the material forming it would be decompressed to reference pressure and temperature.  
    This property is commonly used to study the planets in our Solar System, because differences in $\overline{\rho_0}$ arise only from differences in composition, excluding differences from pressure, temperature or degree of differentiation.  
    In contrast,  the bulk density ($\rho_b$) of planets, which is commonly used in exoplanet studies because of convenience, includes both the effects due to composition, pressure and temperature (the latter not being important for rocky planets). 
    For example, both Mercury and the Earth have a similar $\rho_b$ but different $\overline{\rho_0}$ (see \autoref{fig:rhovsR}). 
    It follows then, that conclusions drawn from comparing the $\rho_b$ of exoplanets \citep{Weiss2014} may pose problems when aiming to study composition.
    
    To avoid this, we introduce the use of uncompressed density for exoplanets and provide a functional fit to our results as to enable anyone to calculate the uncompressed density of a rocky planet given its mass and radius (see \autoref{sec:fit}). 
    We show the results for the planets in our sample  in \autoref{fig:rhovsR}, and include the bulk density for comparison.
    It is clear that in the $\rho_b$ depiction, all super-Earths are denser than Earth, and there is more spread in the values.   
    Instead, the $\overline{\rho_0}$ shows that planets compositions range from $\sim 4-6$ \unit{g/cm^3}.   
    Meaning, it ranges only from Mars' to heavier than Mercury's uncompressed density. 
    Notice though, that the error bars are large.
  
    The higher $\overline{\rho_0}$, the more iron content the planet has. 
    Given the observational biases towards more massive planets, for a given radius, observations are biased towards higher $\overline{\rho_0}$.   
    Thus, the region at low radius and low $\overline{\rho_0}$ is poorly populated due to observational biases. 
    With more super-Earth data we expect this region to become more populated.  
    On the other hand, planets with higher $\overline{\rho_0}$ than Mercury are easier to observe (for a given radius).  
    Intriguingly,  perhaps the most compact planets in the sample, with the highest $\overline{\rho_0}$ are delineating the upper compositional envelope of planet formation.

    Likewise, the region that lacks planets corresponding to $R > 1.65 R_\oplus$ and $\overline{\rho_0}$ above Mercury is truly sparse.
    This is because for planets to have Mercury's composition (or above) with such radius, they need to have a mass over $ 11 M_{\oplus}$.  
    At those masses, all exoplanets lie beyond the RTR except for K2-38b.    
    This planet is particularly puzzling because it has the most iron enrichment of all super-Earths with cmf = ${0.78}^{+0.14}_{-0.21}$, and the largest mass of a rocky planet at $ 12\pm 2.9 M_{\oplus}$ \citep{ref_K2-38b}.
    It is worth observing this planet more to reduce the error uncertainty in both radius and mass. 
    If the iron enrichment proves to be real, we do not have either a compelling planet theory at the moment that can explain such composition \citep{Scora2020}.
    
    In addition, there is a pile-up of planets with a variety of $\overline{\rho_0}$ values at a radius of $\sim 1.5 R_{\oplus}$. 
    This threshold agrees with the radius valley \citep{Fulton2017}, and the suggested value by \citet{Rogers2015} for distinguishing the definitely volatile to possibly rocky planets. 
    With the current data, it is unclear if such diversity only happens in a narrow region of radius, or if it extends to lower radii as well. 
    The combination of the pile-up at $1.5 R_\oplus$ and the real feature of lack of planets with high uncompressed densities above $1.65 R_\oplus$ may be evidence of runaway growth by which planets larger than $1.65 R_\oplus$ acquire enough mass to experience exponential atmospheric accretion and avoid substantial evaporation.
    
    Furthermore,  about $65\%$ of our sample are planets located in the radius gap between $1.5-2\,R_\oplus$ \citep{Fulton2017}, which was observed among planets that only had measured radii.
    \citet{Fulton2017} showed there are mostly two population of small planets: those with a larger radius ($\sim2.4\,R_\oplus$), perhaps volatile planets at lower irradiation values of  $S\sim$ 30 times that of the Earth's ($S_\oplus$), and compact planets ($\sim 1.3R_\oplus$), perhaps rocky ones at higher irradiation values of $S\sim300 S_\oplus$. 
    Importantly, they find much fewer planets in between these sizes and irradiation values, hence the 'radius gap' or 'valley'.
    This feature may indicate that atmospheric evaporation strips planets from their envelopes if irradiation is large enough \citep{Owen2017}.  
    In our sample of planets, 21 have radii between $1.5-2 R_\oplus$ and 9 of these intersect the RTR,  implying they could be volatile. 
    However, at those sizes, the amounts of volatiles would be low.  
    Furthermore, 4 of these RTR-crossing planets are considerably irradiated, at present values above $S>300\,S_\oplus$. 
    This raises an issue, if some of the RTR-crossing planets are volatile, their envelopes would be small, and thus, why were the highly irradiated ones not completely evaporated?   
    Our sample is obviously too small to make categorical conclusions. 
    However, it points to the importance of acquiring good mass and radius data for small exoplanets to test whether or not atmospheric removal is shaping the population of small exoplanets, as well as considering a comparison space where only composition matters (e.g. $\overline{\rho_0}$).
    Along these lines, if indeed the compact planets are the remnants of volatile planets that have suffered evaporation, the planets with high uncompressed densities would suggest that iron enrichment happens early enough in planet formation, while the gas is still around, such that after atmospheric removal the bare iron-rich rocky core is left behind.
    
\subsection{Star-Planets Direct Comparison }
\begin{figure*}
    	\includegraphics[width=\linewidth,height=10cm]{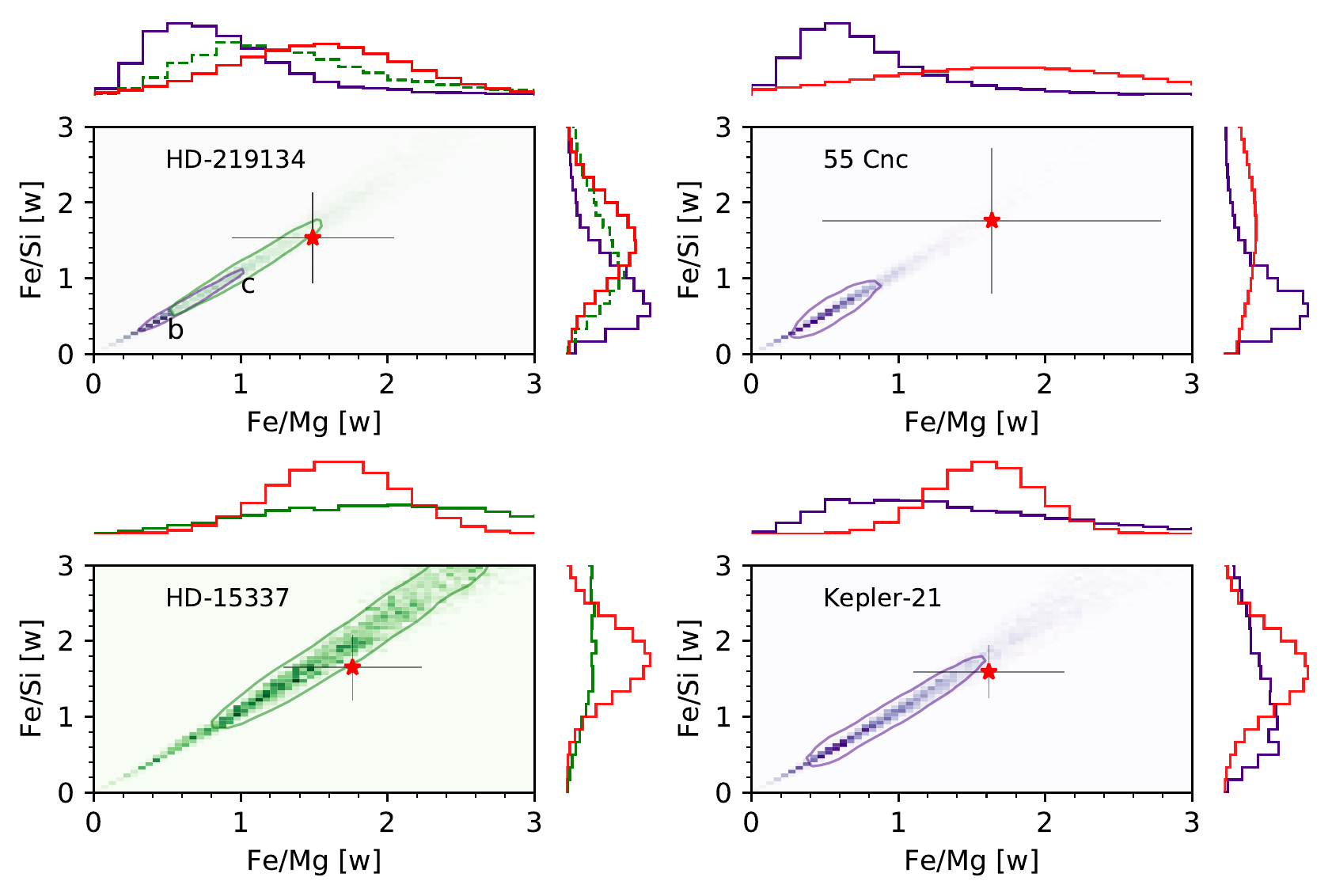}
        \caption{Distribution of chemical refractory ratios for five exoplanets (HD-219134b and c, 55 Cnc e, HD-15337b, Kepler-21b) in comparison to their host star. 
        Purple:  RTR-crossing planets, green: RTR-embedded planets, red: host star}
        \label{fig:St_Pl_2d}
    \end{figure*}

    There are three planetary systems, HD-219134, 55 Cnc, and HD-15337, that have four planets with mass and radius errors below $25\%$, and measured stellar compositions.  
    These systems lend themselves for direct compositional comparisons. 
    In addition, there is a system, Kepler-21, with stellar compositional constraints,  mass and radius measurements for it's planet, but with with a mass error of $34 \%$ \citep{Lopez2016}. 
    We have included this planet in this direct comparison to its host star but excluded it from the population comparison.  
    \autoref{tab:st_pl_chem} and \autoref{fig:St_Pl_2d} shows the star's composition (red) and that of its hosted planets (green/purple).     

\renewcommand{\arraystretch}{1.6} 
\begin{table}
	\centering
	\caption{Stellar and Exoplanet chemical properties}
	\label{tab:st_pl_chem}
	
	\begin{tabular}{lcccc}
        \toprule
		\multicolumn{1}{c}{} &
		\multicolumn{2}{c}{Planet} & \multicolumn{2}{c}{Host Star}\\ 
		\cmidrule(lr){2-3}\cmidrule(lr){4-5}
		Name & Fe/Si & Fe/Mg & Fe/Si & Fe/Mg\\
        \hline
        55 Cnc e & $0.51^{+0.38}_{-0.23}$ & $0.57^{+0.38}_{-0.29}$ & $1.76 \pm 0.96$ & $1.64 \pm 1.15$ \\
        Kepler-21b & $0.94_{-0.53}^{+0.85}$ & $ 0.87_{-0.46}^{+1.0}$& $1.59 \pm 0.35$ & $1.61 \pm 0.52 $\\
        HD-15337b & $2.0^{+1.1}_{-1.1}$ &    $1.9^{+1.5}_{-0.99}$ &  $ 1.65 \pm 0.43$ & $ 1.76 \pm 0.47$\\
        
        HD-219134b &  $0.66^{+ 0.47 }_{-0.32}$&	$0.69^{+ 0.47 }_{-0.35}$ &  $ 1.53 \pm 0.6 $ & $1.49 \pm 0.55$ \\
        HD-219134c &  $1.0^{+0.7}_{-0.44}$ &   $1.2^{+0.55}_{-0.67}$ &  $ 1.53 \pm 0.6 $ & $1.49 \pm 0.55$ \\
		\hline
	\end{tabular}
\end{table}

   \begin{itemize}
\item
    {\bf HD-219134 system}:  The two planets orbiting star HD-219134 seem to have different compositions. 
    Planet c receives 63 times the Earth's received flux, sits well within the rocky region and has very similar refractory ratios to its host star, making it a good candidate for primordial origin. 
    In contrast, Planet b receives 178 times more solar flux than the Earth and intersects the RTR. 
    If this planet is rocky, it would be 2-fold iron-depleted compared to its star, prompting perhaps special formation circumstances. 
    On the other hand, if volatile, a hydrogen-dominated atmosphere would have been susceptible to evaporation, while a water dominated atmosphere would prompt formation scenarios that need to deliver water at these short periods (3 day orbit around a K star). 
    Ruling out the presence of an atmosphere for these two planets within the radius valley, or better yet, obtaining their atmospheres molecular weight, may leap our understanding of how planets form.

    \item
    {\bf 55 Cnc e}:  This enigmatic ultra-hot planet has been observed by multiple groups arriving at different conclusions as to its nature, with some measurements placing it well above the RTR \citep{Demory2011,Winn2011,Endl2012} and more recent work by \citet{ref_55-Cnce} locating it intersecting the RTR.  
    If this planet is considered to be rocky, then its iron content would be depleted with respect to its star. 
    However, the composition of 55 Cnc is poorly known as to support or rule out a primordial composition, and thus, refining the chemical composition of the star is a way to also better understand this planet.

    \item 
    {\bf HD-15337b}:  This warm planet receives 159 times the Earth's insolation flux and is located well within the rocky region. 
    Its mass and radius are consistent with a primordial composition, however, the error bars are too large to make definite conclusions.  
    Improving the data for this planet may indicate there could be a lack of major chemical processing during formation, similar to Earth, at least for some planets with larger masses (i.e. $M_\mathrm{HD-15337b}=7.5 M_\oplus$).

    \item
    {\bf Kepler-21b}:  This planet has the smallest error in radius ($1.16\%$) but has a large error in mass ($34\%$).  
    It intersects the RTR,  and receives a high insolation flux of $ S\sim 2700 S_{\oplus}$.   
    Thus,  similar to HD-219134b, if this planet is volatile, at such high insolation values, the likely candidate for envelope composition is water-dominated, and the problem becomes to explain how a planet with a 2.8 day orbit around a G-type star acquired this much water. 
    If instead, this planet is indeed rocky,  it would be iron-depleted with respect to its host star by a factor of 1.5 in Fe/Mg ratio.  
    However, the error bars in the mass are too large and thus the composition of the planet can also overlap with that of the star.  
    Refining its mass estimate, and ruling out or confirming the presence of an atmosphere can substantially increase our knowledge of this planet's composition and the implication on its formation. 

\end{itemize}

    These examples show the possible pathways to better infer how the composition of the planets is set during formation, given the composition of their host stars. 
    However, the uncertainty in the radius and the mass needs to be improved to obtain meaningful constraints. 

\subsection{Useful Compositional Analytical Fits \label{sec:fit}}
    To facilitate rocky planet interior analysis to other research groups, in this section we present an analytical function $f_\mathrm{cmf}$ that can be used to approximate the core-mass fraction of a rocky planet for a given composition, mass and radius:
    
    \begin{multline}
        \label{eq:cmf}
        f_\mathrm{cmf}(M, R,\mathrm{x}_{\mathrm{Si}}, \mathrm{x}_{\mathrm{Fe}}) = \\ a(\mathrm{x}_\mathrm{Fe}-0.1)+ \sum_{i=0}^1 \alpha_i (\mathrm{x}_{\mathrm{Fe}}-0.1)^i \sum_{j=0}^2 \beta_j \mathrm{x}_{\mathrm{Si}}^j   \\
        \times e^{-bz}\sum_{n=0}^2 \sum_{m=0}^1 c_{nm} \log_{10}\left(\frac{R}{R_\oplus}\right) ^{n} 
        \log_{10}\left(\frac{M}{M_\oplus}\right) ^{m},  
    \end{multline}
    
    where $z$ is the minimum distance between the data point $\left(x_p, y_p \right) = \left( \log_{10}\left(\frac{M}{M_\oplus}\right), \ \log_{10}\left(\frac{R} {R_\oplus}\right) \right) $  and the straight line $y=mx+c$  that corresponds to cmf$=0$ in $\log_{10}$ -space, with $m=0.263157$, $c=0.031716$. 
    Reflected on the $e^{-bz}$ is the fact that our model fits better the compositions that are most similar to Earth.  
    Notice that in the case where the composition is $\mathrm{x_{Si}}=0$ and $\mathrm{x_{Fe}}=0.1$, the summations over $i$ and $j$ becomes 1 (since the indexing starts at zero), reducing the \autoref{eq:cmf} to the last line.

    The equation above can be further used to derive related parameters such as uncompressed density ($\overline{\rho_0}$) or Fe/Si ratio:
    
    \begin{equation}
        \label{eq:rho_0}
        \overline{\rho_0} \approx \frac{\rho_c\rho_m}{ \rho_c + f_\mathrm{cmf}(\rho_m-\rho_c) }\ \ ,\mathrm{\ and}
    \end{equation}
    \begin{equation}
        \label{eq:Fe/Si}
        f_{\mathrm{Fe/Si}} =  \frac{\mu_\mathrm{Fe}}{\mu_\mathrm{Si}} \frac{k_m(0.88-\mathrm{x}_\mathrm{Si})+2k_c \mathrm{x}_\mathrm{Fe}}
        {k_m \mathrm{x}_\mathrm{Si} + k_c(1+\mathrm{x}_\mathrm{py})} \ \ ,
    \end{equation}

    where $\rho_c$ and $\rho_m$ are the reference values for the core and lower mantle composition from the lab experiments (\autoref{Table:EOS_param}) respectively and $\mu_i$ is the atomic weight of one mole of each species.
    The $k_{m}$ and $k_{c}$ are coefficients which depend on the prescribed chemical model in the mantle and core, respectively:
    
    \begin{equation*}
        k_m =  f_\mathrm{cmf}\Big[2(1-\mathrm{x}_\mathrm{Fe})\mu_\mathrm{Mg}+2\mathrm{x}_\mathrm{Fe}\mu_\mathrm{Fe}+(1-\mathrm{x}_\mathrm{py})\mu_\mathrm{Si} + \mu_\mathrm{O}(2\mathrm{x}_\mathrm{py} + 4)\Big]   \  \ ,
    \end{equation*}
    \begin{equation*}
      k_c = (1- f_\mathrm{cmf})\Big[ (0.88-\mathrm{x}_{\mathrm{Si}})\mu_\mathrm{Fe}+0.1 \mu_\mathrm{Ni}+\mathrm{x}_{\mathrm{Si}}\mu_\mathrm{Si}\Big] \  \ ,
    \end{equation*}

    For the purpose of the fit, we created a grid of cmf between 0 and 1 and masses between 1 and 15 $M_\oplus$, which result in radii falling between 1-2 $R_\oplus$.
    
    To obtain the expectation values for the coefficients we performed a MCMC fitting routine by minimizing the log-likelihood function: 
    \begin{equation}
        \mathcal{L} = - \left(\frac{\mathrm{cmf} -f_\mathrm{cmf}(M,R)}{\sqrt{2}\sigma}\right)^2 
    \end{equation}
    where cmf is the actual core-mass fraction and $\sigma$ is the numerical error of the computation.

\begin{figure}
    	\includegraphics[width=\linewidth]{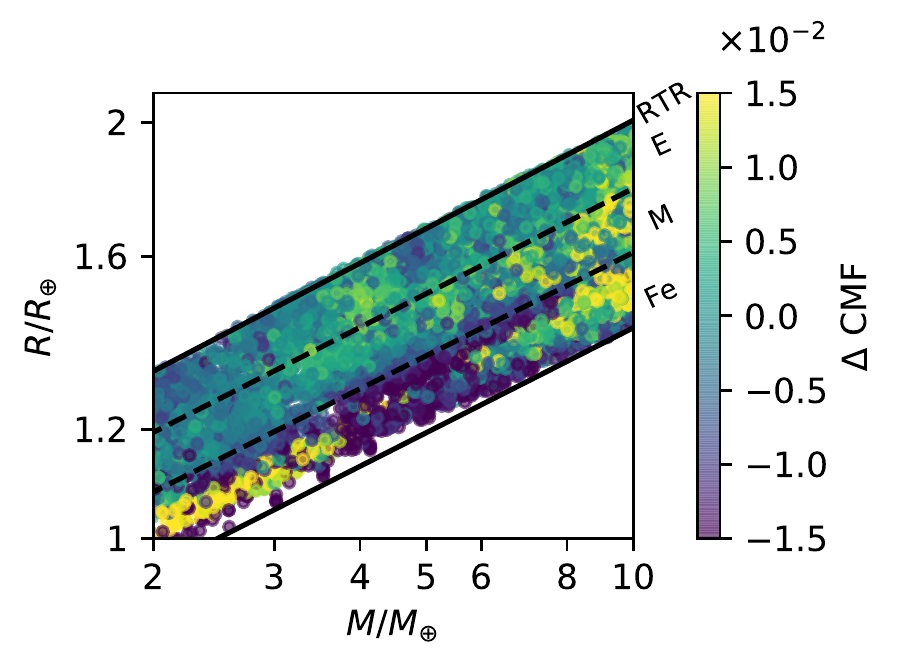}
        \caption{Residuals from our proposed fit to core-mass fraction  as a function of planetary mass and radius (see  \autoref{eq:cmf}). 
        Solid lines represent the  compositions for RTR, Earth (E), Mercury (M) and pure iron (Fe) planets.
        The axes are in logarithmic scale and the colour bar represents the residual value (not as percentage).}
        \label{fig:cmf_fit}
    \end{figure}

    The coefficients obtained are  $a=0.550760$ and $b=2.976432$ , as well as:
    
    \begin{multline*}
       \alpha_i = \begin{pmatrix}
        1 \\
        -0.567941 \\
        \end{pmatrix}
        , \quad 
       \beta_j=
       \begin{pmatrix}
        1 \\
        -3.039501 \\
        1.361339
        \end{pmatrix}
        , \ \\
        c_{nm}=
        \begin{pmatrix}
          0.360172 &   3.071785 \\
          -11.164546 &  -2.225710\\
            6.686932 &  -0.806478
        \end{pmatrix}
        . \hspace{1.cm}
    \end{multline*}
    We show the residuals to our fitted data as differences in cmf in \autoref{fig:cmf_fit}. 
    The agreement is excellent.  
    For example, the fit predicts a cmf of 0.328 for Earth, which corresponds to $\Delta\mathrm{cmf}=0.003$. 
 
    For convenience, we have created an online repository  where the functional fit can be accessed from  this \href{https://github.com/mplotnyko/SuperEarth.py}{URL}.
    
\section{Summary and Conclusions}
    In this study we aimed to compare the composition of rocky planets to those of stars'.
    For this we chose a sample of over 30 planets with measured radius and mass and uncertainty in both below $25\%$.  
    We used a sophisticated internal structure model to constrain the cmf, Fe/Si and Fe/Mg ratios for the planets.  
    This planetary model shows that the most important parameter affecting the radius of a rocky planet is the total amount of Fe (both in the mantle and core) to that of the Mg+Si rocks. 
    Effects from the temperature structure, degree of differentiation, and mantle iron partitioning are small ($\Delta R/R \lesssim 1$).
    
    We compared the planetary chemical ratios to those of planet-hosting stars in a population sense. 
    In this space, the peak of the planets' distribution is at lower Fe/Si values (of 0.9 vs 1.5), and a long tail that extends into compositions of substantial iron-enrichment (Fe/Si$ \gtrsim 5$) compared to stars.  
    This finding calls into question the assumption of using the refractory ratios of the stars as constraints for the planet's composition.
    We distinguished planets that intersect the Rocky Threshold Radius and may be volatile, from those that lie completely within the rocky region.  
    The RTR-crossing planets, if indeed rocky,  would be depleted in iron with respect to the stars by a factor of 2 in their Fe/Si ratios. 
    Without a compelling theory for forming iron depleted planets when compared to their stars, this feature perhaps suggests at least some of those planets are not really rocky.  
    We suggest obtaining phase curves for these planets to rule out the presence of an atmosphere.    

    We also translated the composition of stars, to the equivalent in cmf should a planet be made of the same stellar refractory material. 
    This allowed us to make a comparison in cmf space between stars and planets. 
    We found clearer evidence for core (iron) enrichment in planets compared to stars, with stars distribution dropping at values of cmf=0.5, and planets extending all they way to cmf $\simeq$ 1, (when error bars are considered). 

    In addition, we compared the composition of five planets in four systems directly to the composition of their host stars. 
    In general, we found that the error estimates, especially in mass, preclude us from making definite conclusions. 
    Nevertheless, with the current data, we find interesting questions arising from the compositional comparison. 
    System HD-219134 is particularly intriguing, with planets b and c, both in the radius valley. 
    Planet c appears to have primordial composition whereas inner-planet b appears to be either iron-poor compared to the star, or volatile in nature.
    Either scenario prompts questions about its formation. If the former case is true,  we currently do not have a way to form planets that are depleted in iron. 
    If the latter is true, and the atmosphere is composed mostly of hydrogen, at high insolation values ($\sim180$ that of the Earth's received flux) this atmosphere is susceptible to evaporation; if instead, the atmosphere is water dominated, then we are required to explain how at its short period (3 days) the planet acquired its water. 
    Similar arguments apply to 55 Cnc e and Kepler-21b, because of the very high received flux ( $\gtrsim 2400 S_\oplus$). 
    Although for 55 Cnc e the star's composition is not too well constrained to preclude a primordial composition. 
    Better data in terms of mass, radius and chemical stellar composition will help improve our knowledge of how these planets formed.
    
    Furthermore,  we calculated the uncompressed density and introduced it as a compelling metric to compare composition among planets. 
    We find a real lack of planets above $R\gtrsim 1.65 R_\oplus$ and uncompressed densities $\overline{\rho_0} \gtrsim 5$ \unit{g/cm^3}.  
    This may be pointing to the maximum size a planet can have before it suffers a runaway envelope accretion.  
    On the other hand, planets with sizes near $1.5R_\oplus$ have a wide range of $\overline{\rho_0}$, and it is unclear if this range extends to lower sizes, due to observational biases. 
    Intriguingly, there are a number of planets with higher uncompressed densities than Mercury, near 6 \unit{g/cm^3}. 
    Owing to the fact that for a given radius, planets with higher uncompressed densities are easier to observe, these iron-rich planets may be delineating the maximum iron enrichment attainable in rocky planets. 
    However, the error bars for these planets are large, and thus the inferred iron enrichment needs to be tested with improved mass measurements. 
    In particular, obtaining better mass observations for planets K2-38b, Kepler-105c, Kepler-406b, K2-106b and Kepler-107c, that seem to have large iron contents, may be a direct way to test planet formation theories.

    Lastly, 4 out of 33 planets in our sample have radii smaller than the radius valley ($\lesssim 1.5R_\oplus$), with insolation values above $300 S_\oplus$, and Fe/Mg values that span from approx 0.7 to 5. 
    If these compact planets are indeed a result of atmospheric evaporation, then the iron enrichment and perhaps depletion happens before the gas nebula has dissipated.  
    
    We consider our results to be the first chemical database for low-mass exoplanets that can be used for planetary comparison and our goal is to grow this data set as more planets are observed. 
    Refining the uncertainty, especially in mass, but also radius for planets that appear to be chemical outliers is a compelling way to test formation theories. 
    
    As a final point, with our structure model for rocky planets, we provide fits for core-mass fractions and uncompressed density given planetary mass and radius, for the community to use (\href{https://github.com/mplotnyko/SuperEarth.py}{URL}\footnote{\url{https://github.com/mplotnyko/SuperEarth.py}}).

\section*{Acknowledgements}
    We would like to thank Dan Foreman-Mackey, Francois Soubiran and Gwendolyn Eadie for the useful discussions on the implementation of {\tt EMCEE}, equations of state and statistical methods, respectively, and the reviewer for her/his comments.  MP and DV are supported by the Natural Sciences and Engineering Research Council of Canada (Grant RBPIN-2014-06567) and in part by the National Science Foundation (Grant No. NSF PHY-1748958).
    The research shown here acknowledges the use of the Hypatia Catalog Database\footnote{\url{https://www.hypatiacatalog.com}}, an online compilation of stellar abundance data as described in \citet{Hinkel2014}, which was supported by NASA's Nexus for Exoplanet System Science (NExSS) research coordination network and the Vanderbilt Initiative in Data-Intensive Astrophysics (VIDA).
    This research has made use of the NASA Exoplanet Archive, which is operated by the California Institute of Technology, under contract with the National Aeronautics and Space Administration under the Exoplanet Exploration Program.
    We would like to acknowledge that our work was performed on land traditionally inhabited by the Wendat, the Anishnaabeg, Haudenosaunee, Metis and the Mississaugas of the New Credit First Nation. 
\section*{Data availability}
    The data underlying this article will be shared on reasonable request to the corresponding author.



\bibliographystyle{mnras}
\bibliography{main} 



\appendix
\section{}

In this section we show an example of the results from the MCMC and the interior structure modeling for planet 55 Cnc e, as well as the compositional inferences for all the planets in table form, and in graphical form for the Fe/Si ratio. 

\begin{figure*}
	\includegraphics[width=\linewidth]{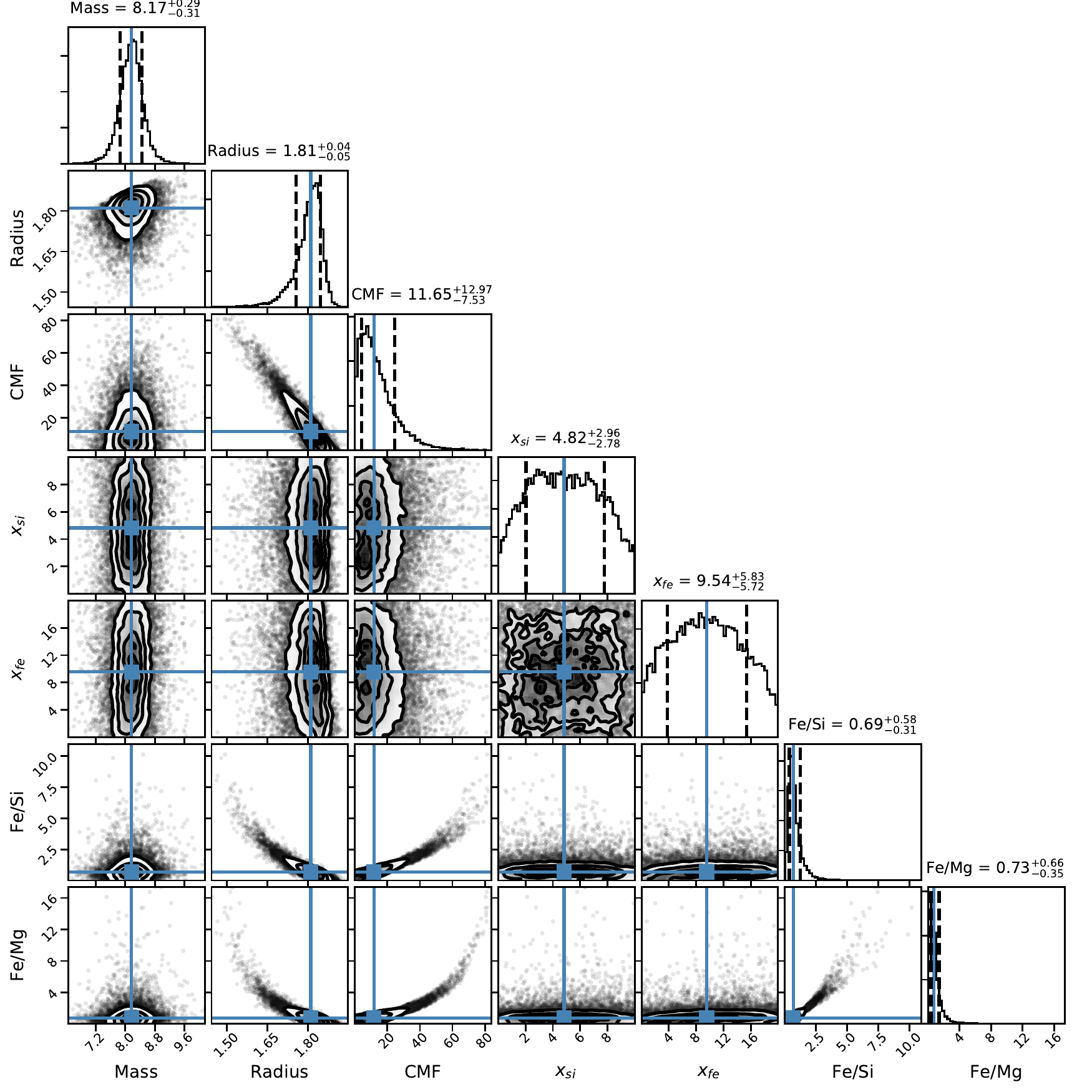}
    \caption{Corner plot result for 55 Cnc e by simulating cmf, amount of iron in the mantle and amount of Si in the core for a given mass and radius. Chemical ratios Fe/Si and Mg/Si are derived quantities.}
    \label{fig:corner}
\end{figure*}  
\begin{figure*}
	\includegraphics[width=\linewidth,height=1.2\linewidth]{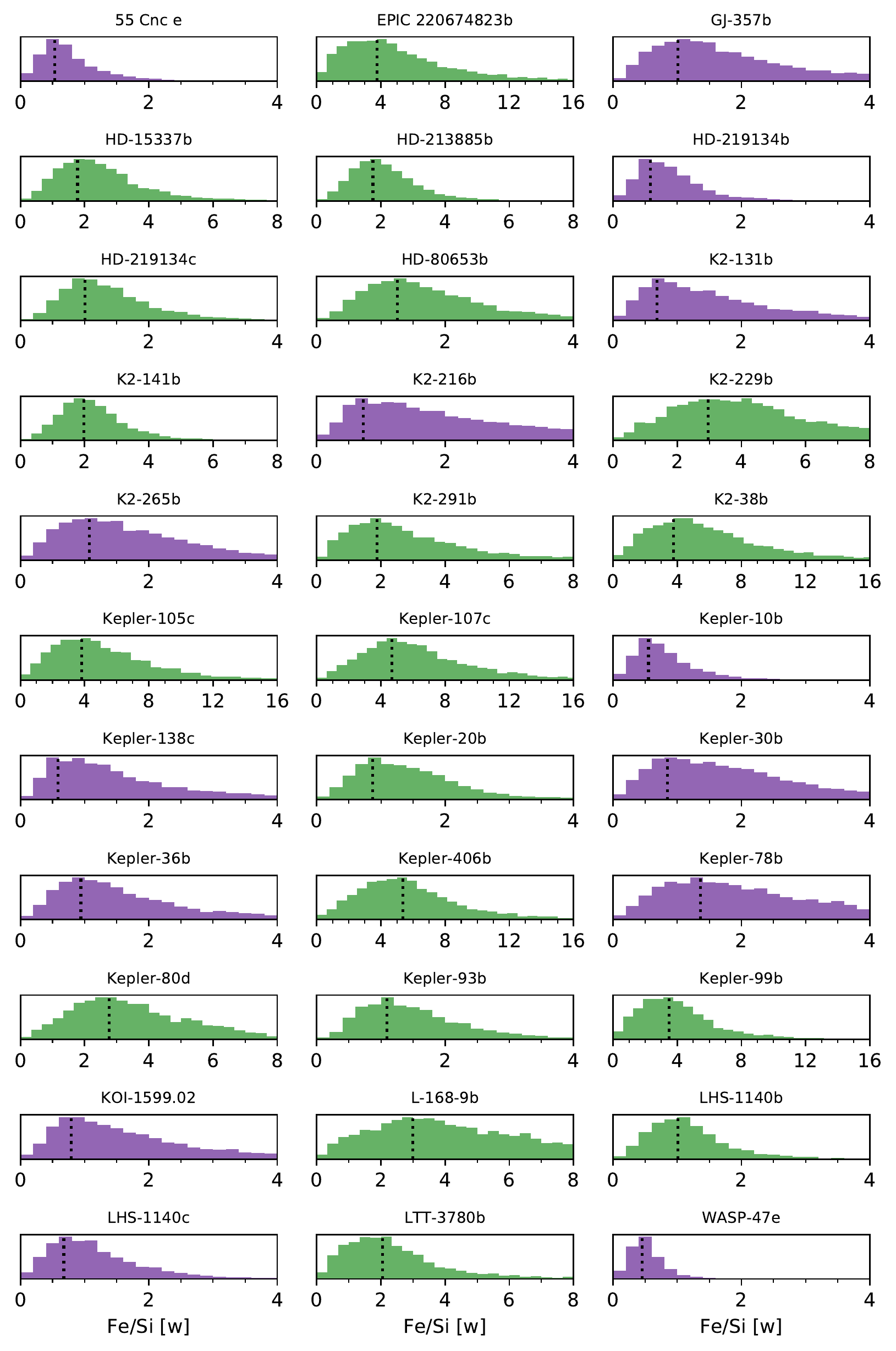}
    \caption{Distribution of Fe/Si ratios obtained from our MCMC simulations for each of the exoplanets with mass and radius uncertainties lower than $25\%$. Purple planets have data that intersect the RTR, whereas green planets do not, and lie completely within the rocky region. Vertical lines show the MAP (or mode) of the planets' distribution. }
    \label{fig:FeSi_sample}
\end{figure*}

\begin{table*}
\caption{Summary of the MCMC results, for rocky planets.}
\label{Table:summary1}
\begin{tabular}{lccccccccl}
\toprule
\multicolumn{2}{c}{RTR-embedded Planets}&
\multicolumn{5}{c}{Parameter}&\multicolumn{2}{c}{$\mathrm{x_{Si}}$ Core and $\mathrm{x_{Fe}}$ Mantle} \\
\cmidrule(lr){1-7}\cmidrule(lr){8-9}

Name &    Mass$^*$ ($M_\oplus$) & Radius$^*$ ($R_\oplus$) &CMF & S$_\oplus$ & $\rho_b$  & $\overline{\rho_0}$ & Fe/Si [w] & Fe/Mg [w]& Reference  \\
  \hline
 {K2-106b$^\dagger$}  &  8.36$^{+0.96}_{-0.94}$ &     1.52$^{+0.16}_{-0.16}$ &   0.68$^{+0.22}_{-0.23}$ &  3500 &  11.0$^{+2.1}_{-2.2}$ &  5.9$^{+1.0}_{-0.9}$ &     2.9$^{+3.1}_{-2.0}$ &     3.4$^{+3.4}_{-2.7}$ &   \citet{ref_EPICb_Guen} \\
       HD-15337b &  7.51$^{+1.09}_{-1.01}$ &     1.64$^{+0.06}_{-0.06}$ &   0.42$^{+0.16}_{-0.16}$ &   160 &   8.7$^{+1.5}_{-1.1}$ &  5.1$^{+0.4}_{-0.5}$ &     2.0$^{+1.1}_{-1.1}$ &    1.9$^{+1.5}_{-0.99}$ &    \citet{ref_HD-15337b} \\
      HD-213885b &  8.83$^{+0.66}_{-0.65}$ &  1.745$^{+0.051}_{-0.052}$ &   0.36$^{+0.11}_{-0.14}$ &  3400 &   9.1$^{+0.8}_{-1.1}$ &  4.9$^{+0.4}_{-0.3}$ &    1.7$^{+1.0}_{-0.73}$ &    1.9$^{+1.2}_{-0.85}$ &   \citet{ref_HD-213885b} \\
      HD-219134c &  4.36$^{+0.22}_{-0.22}$ &  1.511$^{+0.047}_{-0.047}$ &  0.19$^{+0.17}_{-0.09}$ &    63 &   6.8$^{+0.8}_{-0.4}$ &  4.5$^{+0.4}_{-0.2}$ &    1.0$^{+0.7}_{-0.44}$ &   1.2$^{+0.55}_{-0.67}$ &   \citet{ref_HD-219134c} \\
       HD-80653b &   5.6$^{+0.43}_{-0.43}$ &  1.613$^{+0.071}_{-0.071}$ &   0.25$^{+0.17}_{-0.12}$ &  5600 &   7.5$^{+0.9}_{-0.8}$ &  4.6$^{+0.4}_{-0.3}$ &    1.3$^{+0.85}_{-0.7}$ &   1.4$^{+0.91}_{-0.76}$ &    \citet{ref_HD-80653b} \\
         K2-141b &  5.08$^{+0.41}_{-0.41}$ &     1.51$^{+0.05}_{-0.05}$ &   0.42$^{+0.12}_{-0.16}$ &  3300 &   8.1$^{+0.8}_{-0.8}$ &  5.0$^{+0.4}_{-0.4}$ &   1.9$^{+0.96}_{-0.73}$ &    2.1$^{+1.1}_{-0.93}$ &      \citet{ref_K2-141b} \\
         K2-229b &  2.59$^{+0.43}_{-0.43}$ &  1.164$^{+0.066}_{-0.048}$ &    0.6$^{+0.23}_{-0.16}$ &  2500 &   8.4$^{+1.2}_{-1.6}$ &  5.7$^{+0.8}_{-0.7}$ &     3.1$^{+2.5}_{-1.5}$ &     3.3$^{+3.1}_{-1.8}$ &      \citet{ref_K2-229b} \\
         K2-291b &  6.49$^{+1.16}_{-1.16}$ &  1.589$^{+0.095}_{-0.072}$ &   0.39$^{+0.28}_{-0.16}$ &   640 &   8.4$^{+1.8}_{-1.0}$ &  5.0$^{+0.7}_{-0.6}$ &     1.9$^{+1.3}_{-1.2}$ &    1.4$^{+2.1}_{-0.79}$ &      \citet{ref_K2-291b} \\
          K2-38b &    12.0$^{+2.9}_{-2.9}$ &     1.55$^{+0.16}_{-0.16}$ &   0.78$^{+0.14}_{-0.21}$ &   470 &  15.0$^{+0.8}_{-5.0}$ &  6.2$^{+0.9}_{-1.0}$ &     3.8$^{+3.2}_{-2.1}$ &     3.4$^{+4.6}_{-2.0}$ &       \citet{ref_K2-38b} \\
     Kepler-105c &   4.6$^{+0.92}_{-0.85}$ &     1.31$^{+0.07}_{-0.07}$ &   0.72$^{+0.16}_{-0.23}$ &   160 &  11.0$^{+1.7}_{-3.3}$ &  5.8$^{+1.0}_{-0.8}$ &     3.8$^{+2.4}_{-2.0}$ &     3.8$^{+3.6}_{-2.5}$ &  \citet{ref_Kepler-105c} \\
     Kepler-107c &  9.39$^{+1.77}_{-1.77}$ &  1.597$^{+0.026}_{-0.026}$ &   0.67$^{+0.16}_{-0.12}$ &   600 &  12.0$^{+3.0}_{-1.9}$ &  6.0$^{+0.6}_{-0.6}$ &     4.8$^{+2.7}_{-2.2}$ &     5.1$^{+4.0}_{-2.5}$ &  \citet{ref_Kepler-107c} \\
      Kepler-20b &   9.7$^{+1.41}_{-1.44}$ &  1.868$^{+0.066}_{-0.034}$ &   0.28$^{+0.12}_{-0.19}$ &   360 &   8.3$^{+1.5}_{-0.7}$ &  4.5$^{+0.5}_{-0.3}$ &  0.92$^{+0.96}_{-0.41}$ &  0.98$^{+0.99}_{-0.47}$ &   \citet{ref_Kepler-20b} \\
     Kepler-406b &    6.35$^{+1.4}_{-1.4}$ &     1.43$^{+0.03}_{-0.03}$ &   0.73$^{+0.17}_{-0.15}$ &   740 &  11.0$^{+3.0}_{-1.9}$ &  6.1$^{+0.7}_{-0.6}$ &     4.5$^{+2.9}_{-2.2}$ &     4.1$^{+5.7}_{-2.2}$ &  \citet{ref_Kepler-406b} \\
      Kepler-80d &  6.75$^{+0.69}_{-0.51}$ &     1.53$^{+0.09}_{-0.07}$ &   0.57$^{+0.16}_{-0.19}$ &   130 &   9.6$^{+1.8}_{-1.4}$ &  5.4$^{+0.7}_{-0.6}$ &     2.4$^{+1.9}_{-1.1}$ &     2.8$^{+2.6}_{-1.6}$ &   \citet{ref_Kepler-80d} \\
      Kepler-93b &  4.02$^{+0.68}_{-0.68}$ &  1.478$^{+0.019}_{-0.019}$ &   0.22$^{+0.19}_{-0.12}$ &   280 &   6.7$^{+1.0}_{-0.7}$ &  4.5$^{+0.4}_{-0.3}$ &   1.1$^{+0.73}_{-0.55}$ &   1.2$^{+0.91}_{-0.61}$ &   \citet{ref_Kepler-93b} \\
      Kepler-99b &    6.15$^{+1.3}_{-1.3}$ &     1.48$^{+0.08}_{-0.08}$ &   0.62$^{+0.16}_{-0.21}$ &   100 &  11.0$^{+0.6}_{-2.6}$ &  5.5$^{+0.8}_{-0.6}$ &     3.2$^{+2.0}_{-2.0}$ &     2.4$^{+3.5}_{-1.4}$ &   \citet{ref_Kepler-406b} \\
        L 168-9b &   4.6$^{+0.56}_{-0.56}$ &     1.39$^{+0.09}_{-0.09}$ &    0.6$^{+0.22}_{-0.19}$ &   150 &   8.2$^{+2.5}_{-0.5}$ &  5.6$^{+0.7}_{-0.7}$ &     3.1$^{+2.5}_{-1.8}$ &     3.5$^{+2.6}_{-2.4}$ &     \citet{ref_L-168-9b} \\
       LHS 1140b &  6.98$^{+0.89}_{-0.89}$ &  1.727$^{+0.032}_{-0.032}$ &    0.18$^{+0.12}_{-0.10}$ &     0.5 &   7.8$^{+0.5}_{-0.7}$ &  4.5$^{+0.3}_{-0.3}$ &   1.1$^{+0.44}_{-0.58}$ &   1.1$^{+0.58}_{-0.64}$ &    \citet{ref_LHS-1140b} \\
       LTT 3780b &  2.77$^{+0.43}_{-0.43}$ &  1.321$^{+0.074}_{-0.073}$ &      0.40$^{+0.10}_{-0.20}$ &   110 &   6.8$^{+0.9}_{-1.0}$ &  4.8$^{+0.5}_{-0.5}$ &    1.4$^{+1.5}_{-0.85}$ &    1.6$^{+1.5}_{-0.93}$ &    \citet{ref_LTT-3780b} \\
\bottomrule
\end{tabular}
\begin{flushleft}
{$\dagger - $ EPIC 220674823b}\\
$* - $ The radius and mass reported is taken from the observational data. Our simulated distribution may differ for some planets due to the assumption that all planets are rocky (i.e. values below the RTR and above the Fe-Ni compositions are allowed).
\end{flushleft}
\end{table*}

\begin{table*}
\caption{Summary of the MCMC results, for possibly rocky planets.}
\label{Table:summary2}
\begin{tabular}{lccccccccl}
\toprule
\multicolumn{2}{c}{RTR-crossing Planets}&
\multicolumn{5}{c}{Parameter}&\multicolumn{2}{c}{$\mathrm{x_{Si}}$ Core and $
\mathrm{x_{Fe}}$ Mantle} \\
\cmidrule(lr){1-7}\cmidrule(lr){8-9}

Name &    Mass$^*$ ($M_\oplus$) & Radius$^*$ ($R_\oplus$) &CMF & S$_\oplus$ & $\rho_b$ & $\overline{\rho_0}$ & Fe/Si [w] & Fe/Mg [w]& Reference  \\
  \hline
55 Cnc e &  8.08$^{+0.31}_{-0.31}$ &     1.91$^{+0.08}_{-0.08}$ &   0.07$^{+0.11}_{-0.06}$ &  2400 &  7.3$^{+0.5}_{-0.3}$ &  4.2$^{+0.2}_{-0.1}$ &  0.51$^{+0.38}_{-0.23}$ &  0.57$^{+0.38}_{-0.29}$ &            \citet{ref_55-Cnce} \\
     GJ 357b &  1.84$^{+0.31}_{-0.31}$ &  1.217$^{+0.084}_{-0.083}$ &     0.25$^{+0.23}_{-0.16}$ &    13 &  5.9$^{+0.8}_{-0.7}$ &  4.5$^{+0.5}_{-0.4}$ &    1.1$^{+0.88}_{-0.7}$ &   1.0$^{+0.99}_{-0.61}$ &            \citet{ref_GJ-357b} \\
  HD-219134b &  4.74$^{+0.19}_{-0.19}$ &  1.602$^{+0.055}_{-0.055}$ &      0.14$^{+0.10}_{-0.11}$ &   180 &  6.5$^{+0.6}_{-0.3}$ &  4.3$^{+0.2}_{-0.2}$ &  0.66$^{+0.47}_{-0.32}$ &  0.69$^{+0.47}_{-0.35}$ &         \citet{ref_HD-219134c} \\
     K2-131b &     6.5$^{+1.6}_{-1.6}$ &     1.81$^{+0.16}_{-0.12}$ &     0.17$^{+0.24}_{-0.13}$ &  4800 &  7.4$^{+1.3}_{-0.5}$ &  4.4$^{+0.5}_{-0.2}$ &   0.69$^{+1.1}_{-0.35}$ &   0.74$^{+1.0}_{-0.41}$ &        \citet{ref_K2-131b_Dai} \\
     K2-216b &     8.0$^{+1.6}_{-1.6}$ &      1.75$^{+0.17}_{-0.1}$ &     0.25$^{+0.34}_{-0.19}$ &   230 &  7.7$^{+2.1}_{-0.7}$ &  4.5$^{+0.8}_{-0.4}$ &   0.72$^{+1.4}_{-0.32}$ &   0.74$^{+1.5}_{-0.35}$ &            \citet{ref_K2-216b} \\
     K2-265b &  6.54$^{+0.84}_{-0.84}$ &     1.71$^{+0.11}_{-0.11}$ &      0.26$^{+0.20}_{-0.17}$ &   680 &  8.0$^{+1.1}_{-1.2}$ &  4.7$^{+0.5}_{-0.5}$ &   0.98$^{+1.1}_{-0.53}$ &    1.0$^{+1.1}_{-0.55}$ &            \citet{ref_K2-265b} \\
  Kepler-10b &  3.24$^{+0.28}_{-0.28}$ &  1.481$^{+0.049}_{-0.029}$ &     0.09$^{+0.10}_{-0.06}$ &  3600 &  6.0$^{+0.4}_{-0.3}$ &  4.2$^{+0.2}_{-0.2}$ &   0.54$^{+0.35}_{-0.2}$ &  0.57$^{+0.38}_{-0.23}$ &         \citet{ref_Kepler-10b} \\
 Kepler-138c &     5.2$^{+1.2}_{-1.2}$ &     1.67$^{+0.15}_{-0.15}$ &      0.16$^{+0.20}_{-0.12}$ &     5 &  7.1$^{+1.1}_{-0.7}$ &  4.4$^{+0.4}_{-0.3}$ &  0.74$^{+0.82}_{-0.38}$ &  0.77$^{+0.85}_{-0.41}$ &        \citet{ref_Kepler-138c} \\
  Kepler-30b &     8.8$^{+0.6}_{-0.5}$ &        1.9$^{+0.2}_{-0.2}$ &     0.25$^{+0.27}_{-0.18}$ &    22 &  8.6$^{+1.4}_{-1.2}$ &  4.6$^{+0.7}_{-0.4}$ &    0.89$^{+1.4}_{-0.5}$ &   0.74$^{+1.5}_{-0.35}$ &  \citet{ref_Kepler-30b_Hadden} \\
  Kepler-36b &     3.9$^{+0.2}_{-0.2}$ &        1.5$^{+0.1}_{-0.1}$ &      0.21$^{+0.20}_{-0.14}$ &   220 &  6.8$^{+0.9}_{-0.8}$ &  4.5$^{+0.5}_{-0.3}$ &   1.0$^{+0.73}_{-0.58}$ &   1.1$^{+0.76}_{-0.64}$ &         \citet{ref_Kepler-30b_Hadden} \\
  Kepler-78b &  1.87$^{+0.27}_{-0.26}$ &      1.2$^{+0.09}_{-0.09}$ &     0.34$^{+0.23}_{-0.19}$ &  4000 &  6.2$^{+1.2}_{-0.8}$ &  4.6$^{+0.8}_{-0.4}$ &    1.3$^{+1.1}_{-0.76}$ &    1.3$^{+1.3}_{-0.73}$ &    \citet{ref_Kepler-78b_Grun} \\
 KOI-1599.02 &     9.0$^{+0.3}_{-0.3}$ &        1.9$^{+0.2}_{-0.2}$ &     0.21$^{+0.23}_{-0.16}$ &    78 &  8.3$^{+1.1}_{-1.0}$ &  4.5$^{+0.5}_{-0.3}$ &  0.83$^{+0.99}_{-0.41}$ &  0.89$^{+0.96}_{-0.47}$ &        \citet{ref_KOI-1599.02} \\
   LHS 1140c &  1.81$^{+0.39}_{-0.39}$ &  1.282$^{+0.024}_{-0.024}$ &    0.11$^{+0.21}_{-0.09}$ &     6 &  5.7$^{+0.5}_{-0.6}$ &  4.3$^{+0.4}_{-0.2}$ &   0.66$^{+0.7}_{-0.32}$ &  0.74$^{+0.76}_{-0.41}$ &          \citet{ref_LHS-1140b} \\
    WASP-47e &  6.83$^{+0.66}_{-0.66}$ &   1.81$^{+0.027}_{-0.027}$ &  0.05$^{+0.06}_{-0.04}$ &  3900 &  7.1$^{+0.3}_{-0.3}$ &  4.2$^{+0.1}_{-0.1}$ &    0.45$^{+0.2}_{-0.2}$ &   0.45$^{+0.23}_{-0.2}$ &           \citet{ref_WASP-47e} \\

\bottomrule
\end{tabular}
\begin{flushleft}
$* - $ The radius and mass reported is taken from the observational data. Our simulated distribution may differ for some planets due to the assumption that all planets are rocky (i.e. values below the RTR and above the Fe-Ni compositions are allowed).
\end{flushleft}
\end{table*}


\bsp	
\label{lastpage}
\end{document}